\documentclass[aps,prb,twocolumn,superscriptaddress,usenames,dvipsnames,longbibliography]{revtex4-2}
\usepackage{hyperref}
\usepackage{bm}
\usepackage[T1]{fontenc}
\usepackage{amsmath}
\usepackage{graphicx}
\usepackage[dvipsnames]{xcolor}
\usepackage{tabularx}
\usepackage{multirow}
\usepackage{colortbl}
\usepackage{soul}
\usepackage{cancel}

\usepackage{xr}
\usepackage[normalem]{ulem}
\hypersetup{colorlinks=true, linkcolor=blue, citecolor=red, urlcolor=magenta, pdftitle={Large anti-magnetoelectricity from spin and orbitals in ferroelectric BiCoO3}, pdfauthor={M. Braun, B. Guster, E. Bousquet}}\usepackage{natbib}
\definecolor{gray}{rgb}{0.8,0.8,0.8}

\graphicspath{ {./IMG/} }

\def\bco	{{BiCoO$_3$}}
\def\s2s21  {{$\sqrt{2}\times\sqrt{2}\times1$}}
\def\ab     {{\textit{ab initio}}}

\newcommand{\maxime}[1]{\textcolor{red}{{#1}}}

\setstcolor{red}

\begin{document}
\title{Large dynamical magnetic effective charges and anti-magnetoelectricity from spin and orbital origin in  multiferroic \texorpdfstring{BiCoO$_3$}{}
}


\author{Maxime Braun}
\affiliation{Physique Th\'eorique des Mat\'eriaux, QMAT, CESAM, Universit\'e de Li\`ege, B-4000 Sart-Tilman, Belgium}
\affiliation{Univ. Lille, CNRS, Centrale Lille, ENSCL, Univ. Artois, UMR 8181-UCCS-Unité de Catalyse et Chimie du Solide, F-59000 Lille, France}
\author{Bogdan Guster}
\affiliation{Physique Th\'eorique des Mat\'eriaux, QMAT, CESAM, Universit\'e de Li\`ege, B-4000 Sart-Tilman, Belgium}
\author{Andrea Urru}
\affiliation{Department of Physics and Astronomy, Rutgers University, Piscataway, New Jersey 08854, USA}
\author{Houria Kabbour}

\affiliation{Univ. Lille, CNRS, Centrale Lille, ENSCL, Univ. Artois, UMR 8181-UCCS-Unité de Catalyse et Chimie du Solide, F-59000 Lille, France}

\author{Eric Bousquet}
\affiliation{Physique Th\'eorique des Mat\'eriaux, QMAT, CESAM, Universit\'e de Li\`ege, B-4000 Sart-Tilman, Belgium}

\begin{abstract}
Using first-principles calculations, we explore the magnetoelectric properties of the room-temperature multiferroic crystal \bco.
We use both applied magnetic field and finite-difference techniques to show that \bco~is anti-magnetoelectric at the linear level.
The calculation of the dynamical effective charges reveals that the total magnetoelectric response is zero due to the compensating non-zero magnetoelectric response of each magnetic sublattice.
This calculation also highlights that the the orbital contribution to the response is remarkably larger than the spin one and that each sublattice has a rather large total magnetoelectric response of 85 ps/m.
Furthermore, we provide an intuitive recipe to visualize the dynamical magnetic effective charge, allowing to examine its multipolar nature which we confirm by means of \textit{ab initio} calculations.
Given the large value of the local response, we investigate the ferromagnetic phase as well, which gives a giant magnetoelectric response of about 1000 ps/m and coming mainly from the spin contribution this time.
Finally, we discuss the possible reasons for such a large magnetoelectric response in \bco~and propose possible strategies to unveil this potentially large response.
\end{abstract}

\maketitle

\section{Introduction}
\label{sec:intro}

Magnetoelectricity, that is the coupling between the electric and magnetic orders in materials, experienced an increased interest in the past 20 years \cite{Fiebig2005,Eerenstein2006,Rivera2009} due to its promising technological applications~\cite{Kosub2017, Kopyl2021, Liang2021,Spaldin2019}. 
Although the idea of a magnetoelectric (ME) effect was first proposed by Röntgen in 1888 \cite{Roentgen1888}, with Pierre Curie having emphasizing in 1894 by symmetry analysis that intrinsic MEs could exist \cite{Curie1894}; the name magnetoelectricity was coined only a few decades later by Debye \cite{Debye1926}. Cr$_2$O$_3$ was the first single crystal to be theoretically proposed \cite{Dzyaloshinski1960} and, soon after, experimentally demonstrated \cite{Astrov1960,Astrov1961,Folen1961,Rado1961} to show a ME effect.

ME effects show up as cross-coupling terms in a Landau free energy expansion in terms of applied magnetic ($\boldsymbol{\mathcal{H}}$) and electric ($\boldsymbol{\mathcal{E}}$) fields:

\begin{equation}
    F_{\text{ME}} = -\alpha_{ij} \mathcal{E}_{i} \mathcal{H}_{j} - \frac{1}{2} \beta_{ijk}\mathcal{E}_{i} \mathcal{E}_{j} \mathcal{H}_{k} - \frac{1}{2} \gamma_{ijk}\mathcal{H}_{i} \mathcal{H}_{j} \mathcal{E}_{k} + \dots .
    \label{eq:ME}
\end{equation}

\noindent
Particularly, here the first term, identified by the second-rank tensor $\alpha_{ij}$ defined as
\begin{equation}
    \alpha_{ij}=\mu_{0}\frac{\delta M_{j}}{\delta \mathcal{E}_{i}}\biggr\rvert_{\mathcal{H}} =\frac{\delta P_{i}}{\delta \mathcal{H}_{j}}\biggr\rvert_{\mathcal{E}},
\end{equation}
\noindent 
describes the so-called \textit{linear} ME effect, which results in a net magnetization $\mathbf{M}$ (polarization $\mathbf{P}$) linear in the applied electric (magnetic) field's amplitude. 
On the other hand, $\beta_{ijk}$ and $\gamma_{ijk}$ in Eq.~\eqref{eq:ME} are third-rank tensors accounting for the (non-linear) quadratic ME coupling.
Because of the following upper bound for the ME response~\cite{Brown1968}:
\begin{equation}\label{eq:upperbond}
    \alpha_{ij}^{2} \leq \epsilon_{0} \mu_{0} \epsilon_{ii} \mu_{jj},
\end{equation}
where $\epsilon_0$, $\mu_0$, $\epsilon_{ii}$ and $\mu_{jj}$ are the vacuum and material permittivities and permeabilities respectively, the ME coupling is expected to be large in multiferroic materials due to the combined large electric permittivity and large magnetic permeability~\cite{bayaraa2021,Sasani2022,fernandez2016, bousquet2011a, dieguez2011,chun2010}. 
For this reason a great effort has been made to understand the different mechanisms of magnetic and ME interactions in multiferroics from theory,  simulations and experiments~\cite{malashevich2012,Iniguez2008,birol2012,Bousquet2016,Fiebig2016,Xu2024}.

On the computational side, it is convenient to split the full ME response into electronic, lattice-mediated and strain-mediated contributions $\alpha^{tot} = \alpha^{\text{elec}} + \alpha^{\text{latt}} + \alpha^{\text{str}}$. 
Furthermore, each term is usually separated into the contributions coming from the spin and orbital degrees of freedom.
Such calculations have shown that the lattice mediated terms can have a significant contribution to the full response when close to a ferroelectric phase transition where a polar soft mode is at play~\cite{Iniguez2008,Ye2014, bousquet2011a} and that the electronic contribution is large when close to a metal-insulator phase transition~\cite{bousquet2011a}. 
More importantly, when the quenching of the orbital angular momentum is absent, a strong contribution from the orbital magnetization can also be at play~\cite{Scaramucci2012,solovyev2016,malashevich2012,yanagi2018}. 
For example, this is the case for LiFePO$_4$ where the lattice-mediated orbital contribution to $\alpha_{\parallel}$ accounts almost entirely for the in-plane ME response at T$=$0 K \cite{Scaramucci2012}.

Over the past couple of decades, an increasing interest to a convenient alternative description of linear ME responses  based on microscopic markers named ME multipoles has been observed in the literature~\cite{ederer2007, spaldin2013, Urru2022, Verbeek2023}. 
These markers identify the second-order contribution in a multipole expansion of the energy of a magnetization density $\boldsymbol{\mu} (\mathbf{r})$ interacting with a non-uniform magnetic field $\boldsymbol{\mathcal{H}} (\mathbf{r})$: 
\begin{eqnarray}
E & = & -  \int \boldsymbol{\mu}(\mathbf{r}) \cdot
\boldsymbol{\mathcal{H}}(\mathbf{r}) \, d^3 \mathbf{r} \nonumber \\
 & = & - \boldsymbol{\mathcal{H}}(0) \cdot \int \boldsymbol{\mu}(\mathbf{r}) \, d^3 \mathbf{r} 
- \partial_{i} \mathcal{H}_{j}(0) \int r_{i} \mu_{j} (\mathbf{r}) \, d^3 \mathbf{r} \quad  - \ldots \nonumber \\
\label{eq:dipole}
\end{eqnarray}
The tensor $\mathcal{M}_{ij} = \int r_{i} \mu_{j} (\mathbf{r})$ appearing in Eq.~\eqref{eq:dipole}, is called ME multipole tensor and has a one-to-one link to the linear ME response $\alpha_{ij}$, because they both break separately spatial- ($\mathcal{I}$) and time-inversion ($\mathcal{T}$) symmetries, while preserving the combined symmetry $\mathcal{TI}$.

ME multipoles can be defined both for the whole unit cell or for single atoms, depending on whether the integration in Eq.\ \eqref{eq:dipole} is performed over the unit cell or over a sphere centered at the atomic site, respectively~\cite{spaldin2013}. 
For the purpose of the present work, we will only refer to the latter, i.e., the so-called \textit{atomic-site multipoles}, which identify the contribution to the ME response by each single atom. 
Practically, a ferroic arrangement of a multipole appearing in given entries $\mathcal{M}_{ij}$ of the multipole tensor results in corresponding non-zero entries $\alpha_{ij}$ of the linear ME tensor. 
For example, ferroically ordered ME monopoles $a$ identify a net diagonal and isotropic linear ME response.

Very recently, it has been shown that a vanishing net ME response does not prevent a local, atomic ME response~\cite{Verbeek2023}. 
In fact, there can be cases where the local ME responses of the oppositely aligned antiferromagnetic sublattices are opposite to each other, thus giving a zero bulk net response~\cite{Verbeek2023}. 
Such effect, named \textit{anti-ME}, corresponds microscopically to antiferroically ordered ME multipoles.

In this work, we explore the anti-ME effect in the multiferroic material BiCoO$_3$ by means of \ab~ density functional theory (DFT). 
We specifically consider the electronic and lattice contributions to the ME coupling in the C-type antiferromagnetic (C-AFM) and ferromagnetic (FM) phases of \bco. 
This work is organized as follows: 
in Sec. \ref{sec:technical} we summarize the theoretical approach and list the computational parameters for each code used in our calculations; 
in Sec. \ref{sec:struc_lin_resp} we describe the structural and magnetic properties of \bco; 
in Sec. \ref{sec:elec_mag_latt_resp} we focus on the Born effective charges and dynamical magnetic effective charges and their contribution to the ME tensor;
in Sec. \ref{sec:me_response} we rationalize the absence of ME response in the C-AFM ground-state state of \bco~and the potentially large ME response of the FM phase; 
in Sec. \ref{sec:conclusions} we give a conclusion and we provide some suggestions on how the large ME response of the FM phase could be unlocked in \bco.

\section{Methods}
\label{sec:technical}

\subsection{Theoretical formalism}
\label{sec:theory_ME}

\subsubsection{Electronic and  lattice-mediated ME responses} The electronic (i.e., clamped-ions) contribution to the ME tensor was evaluated by applying a Zeeman magnetic field and computing the induced polarization, without further relaxing the ions. 
In this approach, the polarization $\mathbf{P}$ induced by a magnetic field $\boldsymbol{\mathcal{H}}$ is
\begin{equation}
    P_i (\boldsymbol{\mathcal{H}}) = \frac{1}{\Omega} \alpha_{ij}^{\text{elec}} \mathcal{H}_{j},
    \label{eq:alpha_elec}
\end{equation}
where $\alpha_{ij}^{\text{elec}}$ is the electronic contribution to the ME response, and $\Omega$ is the unit-cell volume.

The lattice-mediated contribution to $\alpha$ can be obtained in principle within the same Zeeman field approach by further relaxing the atomic positions in presence of the field. 
However, we note that in this way the orbital contribution to $\alpha_{ij}^{\text{elec}}$ would not be accounted for because the Zeeman field acts only on the spin degrees of freedom. 
Thus, we used a method related to the one proposed in Ref.~\cite{Iniguez2008} where we estimated the lattice-mediated part of the ME response as follows:
\begin{equation}
    \alpha^{\text{latt}}_{ij} = \frac{\mu_{0}}{\Omega} \sum_{n=1}^{N_{\text{IR}}} \frac{ S_{n,ij} }{\omega_{n}^2},
    \label{eq:alpha_latt_1}
\end{equation}
\noindent
with $\omega_n$ the eigenvalues of the dynamical matrix (i.e., phonon \textit{angular} frequencies 
\footnote{Note that the phonon angular frequencies $\omega$ are related to the frequencies $\nu$, commonly expressed in THz or cm$^{-1}$, by $\omega = 2 \pi \nu$.}) 
corresponding to the $N_{\text{IR}}$ infrared-active phonon modes, whereas  $S_{n,ij}$ is, what we will call the ME mode-oscillator strength tensor, defined as follows:
\begin{equation}
S_{n,ij}=\left(\sum_{\kappa,i'}Z^{*\text{m}}_{\kappa,ii'}{u}_{n,\kappa,i'}\right)\times\left(\sum_{\kappa,j'}Z^{*\text{e}}_{\kappa,j'j}{u}_{n,\kappa,j'}\right),
    \label{ME_Mode-oscillator}
\end{equation}
where $\kappa$ identifies the atom in the unit cell,  ${u}_{n,\kappa,j}$ the normalized eigendisplacement, i.e., the generalized eigenvector of the interatomic force constants matrix (see Supplementary Material for a more detailed discussion), $Z_\kappa^{*\text{e}}$ and $Z_\kappa^{*\text{m}}$  the Born effective charge (BEC) and dynamical magnetic effective charge (DMC), respectively, defined as:
\begin{equation}
    \label{eq:becs}
    Z^{*\text{e}}_{\kappa,ij} = \Omega \frac{\partial P_{i}}{\partial \tau_{\kappa,j}} =\frac{\partial^2E}{\partial\mathcal{E}_{i}\partial \tau_{\kappa,j}},
\end{equation}
\noindent 
i.e., the net polarization along direction $i$ induced by a displacement $\tau$ of atom $\kappa$ along direction $j$, and
\begin{equation}
    \label{eq:dmcs}
    Z^{*\text{m}}_{\kappa,ij} = \Omega \frac{\partial M_{i}}{\partial \tau_{\kappa,j}}=\frac{\partial^2E}{\partial\mathcal{H}_{i}\partial \tau_{\kappa,j}},
\end{equation}
\noindent 
which corresponds to the net magnetization along direction $i$, induced by a displacement $\tau$ of atom $\kappa$ along direction $j$~\cite{Ye2014}.
Hence, this ME mode oscillator strength formula is similar to the so-called mode oscillator strength of polar modes~\cite{Gonze1997} but here for modes that are both infrared active and magnetically active (i.e. the phonon mode carries a net magnetization).

We note that the scheme proposed by I\~nigu\'ez in Ref.~\cite{Iniguez2008} to compute the lattice-mediated ME response is based on the so-called mode effective charges and mode dynamical magnetization, $\bar{Z}^{*\text{e}}_{n}$ and $\bar{Z}^{*\text{m}}_{n}$ respectively, of the $n$-th phonon mode. 
These identify the net polarization and net magnetization linearly induced  by a phonon mode, and are expressed as follows:
\begin{equation}
    \Bar{Z}^{*\text{e}}_{n,i} = \sum_{\kappa,j} Z^{*\text{e}}_{\kappa,ij} \Tilde{u}_{n,\kappa,j},
    \label{eq:zbare}
\end{equation}
\begin{equation}
    \Bar{Z}^{*\text{m}}_{n,i} = \sum_{\kappa,j} Z^{*\text{m}}_{\kappa,ij} \Tilde{u}_{n,\kappa,j}.
    \label{eq:zbarm}
\end{equation}
It is worthwhile noting that the quantities identified by Eqs. \eqref{eq:zbare} and \eqref{eq:zbarm} are different from those inside the round brackets in Eq. \eqref{ME_Mode-oscillator}, because the eigendisplacements $u_{n, \kappa, j}$ and $\tilde{u}_{n, \kappa, j}$ are normalized differently. 
We refer the reader to the Supplementary Material for a more detailed discussion.

\subsubsection{ME multipoles}

The ME multipole tensor $\mathcal{M}$ introduced above is usually conveniently decomposed into its spherical components \cite{spaldin2013}, namely the ME monopole
\begin{equation}
a =  \frac{1}{3} {\cal M}_{ii} = \frac{1}{3} \int \mathbf{r} \! \cdot \boldsymbol{\mu}(\mathbf{r}) \, d^3 \mathbf{r},
\end{equation}
the toroidal moment vector
\begin{equation}
    \mathbf{t} = \frac{1}{2}  \int \mathbf{r} \! \times \boldsymbol{\mu}(\mathbf{r}) \, d^3 \mathbf{r},
\end{equation}
having components
\begin{equation}
    t_i = \frac{1}{2} \varepsilon_{ijk} {\cal M}_{jk} \quad ,
\end{equation}
and the traceless quadrupole tensor
\begin{eqnarray}
    q_{ij} &=& \frac{1}{2}\left({\cal M}_{ij} + {\cal M}_{ji} - \frac{2}{3} \delta_{ij} {\cal
M}_{kk}\right)\nonumber\\
&= &\frac{1}{2} \int \left[r_i \mu_j (\mathbf{r}) +
r_j \mu_i (\mathbf{r}) - \frac{2}{3} \delta_{ij} \mathbf{r}\! \cdot \boldsymbol{\mu}(\mathbf{r})
\right] d^3\mathbf{r},
\label{eq:quadrupole}
\end{eqnarray}
such that the overall ${\cal M}$ tensor is:
\begin{equation}
\label{eq_me_tensor}
\begin{pmatrix}
a + \frac{1}{2}q_{x^2-y^2} - \frac{1}{2} q_{z^2} 
& t_z + q_{xy} & t_y + q_{xz}\\
-t_z + q_{xy} & a - \frac{1}{2}q_{x^2-y^2} - \frac{1}{2} q_{z^2}  & -t_x + q_{yz} \\
-t_y + q_{xz} & t_x + q_{yz} &  a + q_{z^2}
\end{pmatrix}
.
\end{equation}
As stated earlier, ferroically ordered ME multipoles appearing in a given entry $ij$ of the multipole tensor $\mathcal{M}$ result in a net ME response $\alpha_{ij}$. 
For example, ferroically ordered monopoles $a$ imply a net diagonal and isotropic ME response.

\subsection{\textit{Ab initio} calculations}
First-principles calculations were performed within DFT and density functional perturbation theory (DFPT)~\cite{Gajdo2006}, as implemented in the Vienna Ab Initio Simulation package (VASP 5.4.4)~\cite{Kresse1996,Kresse1993} and ABINIT (9.11.0)~\cite{Gonze2020} codes. 
We crosschecked between VASP and ABINIT codes to compute the lattice-mediated contribution to the ME response (Eq. \eqref{eq:alpha_latt_1}). 
Further on, we used ABINIT to compute the electronic contribution to the ME response (Eq.~\eqref{eq:alpha_elec}). 
In both codes, phonon eigenvalues and eigenvectors at zone center and BECs were computed using DFPT, while DMCs were computed using a finite differences (FD) approach. 
The calculation of orbital magnetization was carried
out using the muffin-tin approximation by summing the partial atomic contributions in the PAW spheres and neglecting the contribution in the interstitial region.
We note that to get a complete evaluation of the orbital magnetization, including the interstitial contributions, the modern theory of magnetization would be required ~\cite{Resta2010,Ceresoli2010,THONHAUSER2011}. 

Next, we present the computational parameters used for both codes.

\subsubsection{VASP}
Lattice relaxation calculations were done within the generalized gradient approximation (GGA) with a revised exchange–correlation functional for solids (PBEsol) \cite{Perdew2009} and a relaxation threshold of $10^{-8}$ eV/\AA \, on the ionic forces was used. 
Electrons were described using Projector Augmented-Wave (PAW) pseudopotentials, with the following valence electrons configurations: Bi $6s^26p^3$ (dataset \texttt{Bi}), Co $4s^13d^8$ (dataset \texttt{Co\_sv}), O $2s^22p^4$ (dataset \texttt{O}). 
The pseudo-wave functions were expanded into plane waves \cite{Blöchl1994} with a kinetic energy cut-off of 650 eV. 
The Brillouin zone (BZ) of the magnetic unit cell, corresponding to a $\sqrt{2}$ ×$\sqrt{2}$×$1$ crystal supercell (see Section \ref{sec:struc_lin_resp}, was sampled by a 12$\times$12$\times$12 Monkhorst-Pack k-point mesh \cite{Monkhorst-prb1976}. 

DMCs were computed with the FD technique explained above, with self-consistent field (SCF) calculations performed at the distorted structures obtained by moving the atoms with a set of finite displacements of $\pm$ 0.01, 0.03, and 0.05 \text{\r{A}} for both the in-plane and out-of-plane responses.
The non-collinear SCF calculations with spin-orbit coupling (SOC) to compute the DMCs were performed within the local-density approximation (LDA), using into account the Slater exchange and Perdew-Zunger parametrization of Ceperley-Alder Monte-Carlo correlation data \cite{Perdew1981}. 
We used  a strict convergence threshold of $10^{-12}$ eV of the total energy on the self-consistent cycle and added 500-2000 steps to allow for the convergence of the magnetic moment in order to determine the DMCs accurately. 
Correlation effects were dealt with by applying a Hubbard correction~\cite{Liechtenstein1995} to the d orbitals of Cobalt, with a U parameter of 6 eV, which opens a band gap (1.42 eV) in agreement with previously reported theoretical values \cite{Ravindran2008,Uratani2005} and comparable to experimental values from X-ray emission and absorption spectroscopy (XES/XAS) \cite{McLeod2010} (1.7 eV).

The ME and higher-order multipoles were computed by decomposing the DFT+U occupation matrix into spherical tensor moments \cite{nordstrom2009}, as obtained from an extension of the Multipyles Python package \footnote{See \url{https://github.com/materialstheory/multipyles}.}.
\paragraph*{Visualization of DMC-related perturbed magnetization densities.}
In order to create an isosurface to illustrate the DMCs, we calculated at each point of the $96 \times 96 \times 80$ Fast Fourier Transformation (FFT) grid the difference in magnetization density  between the perturbed and ground states.
The display limit value, called the isosurface level, was set in order to get a similar surface area for both the linear and quadratic contributions shown in Fig.~\ref{DecompCobaltDMC}, discussed below in Sec.~\ref{dmc_sec}, for visual clarity. 
In order to decompose the total response into linear and  quadratic contributions (Fig.~\ref{DecompCobaltDMC}), 
we first worked out the average of the magnetization density from two opposite atomic displacements, which allowed us to get the quadratic part. 
Then, we subtracted it from the total induced magnetization density to get the linear contribution.

\subsubsection{ABINIT}

Calculations with ABINIT were performed in the LSDA+U framework~\cite{Torrent2008} using the JTH PAW atomic potentials (v1.1)~\cite{Jollet2014}, with SOC included and starting from the relaxed structures obtained in VASP. 
A Hubbard-U correction of 5.0 eV was applied to the 3d Co shell, while although different from the one used in VASP, it gives  the same band gap (since the effective U is affected by the radii of the PAW spheres, which are different in the two codes). 
The Kohn-Sham pseudo-wave functions were expanded over a plane-wave basis set with a kinetic energy cut-off of 20 Ha (544 eV). 
We used a strict convergence threshold of $2.7\times10^{-12}$ eV for the total energy on the self-consistent cycle, to determine the DMCs accurately. 
The BZ of the magnetic unit cell was sampled with a Monkhorst-Pack k-point grid of 8x8x8  points. 
For DFPT calculations~\cite{Gonze1997}, we fully re-optimized the structure, by using a tolerance of 1.3$\times 10^{-6}$ GPa on the internal pressure and of 3$\times 10^{-7}$ eV/\text{\r{A}} on the forces. 


\section{Structural and magnetic properties}
\label{sec:struc_lin_resp}

\bco~crystallizes in a supertetragonal phase described by the space group $P4mm$ (point group 4mm) with a much more pronounced tetragonality ($c/a$ = 1.267) than the reference PbTiO$_3$ crystal for this phase. 
The crystal unit cell contains 5 atoms, the Co atoms are situated in a tetragonal prism having four O atoms at the corner of the base and one at the apex, side-sharing the base with an opposing trigonal prism with the Bi atom at the apex.
Particularly, Bi atoms sit at the 1a (0, 0, 0) Wyckoff site, Co atoms are located at the 1b Wyckoff position ($\frac{1}{2}$, $\frac{1}{2}$, z$_{Co}$), whereas the apical and basal oxygens are found at the 1b ($\frac{1}{2}$, $\frac{1}{2}$, z$_{O_{api}}$) and 2c ($\frac{1}{2}$, 0, z$_{O_{bas}}$), respectively.

\begin{figure}[t]
\centering
\includegraphics[width=0.95\linewidth]{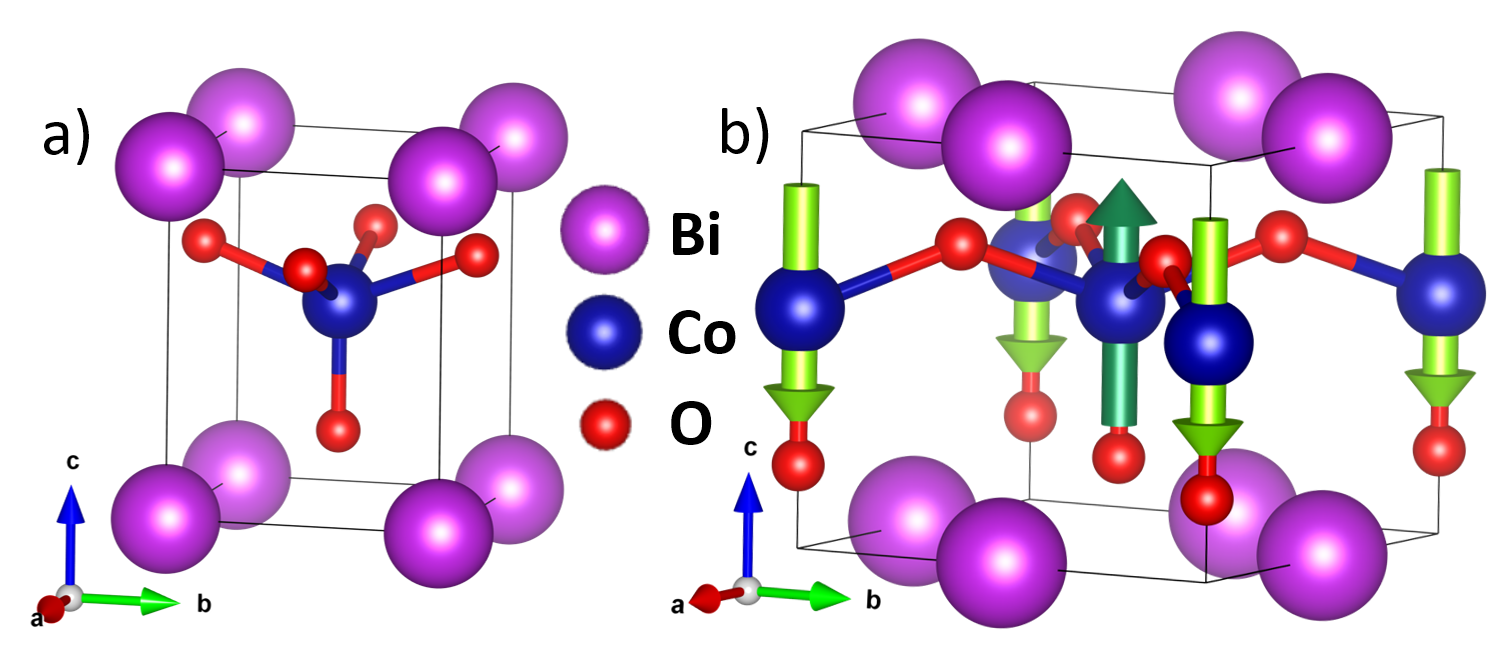}
\caption{Schematic representation of a) the  5-atoms primitive cell  of BiCoO$_3$ without magnetism  and b) the $\sqrt{2}$×$\sqrt{2}$×1 C-AFM unit cell. Green arrows show the direction of the Co magnetic moment.}
\label{fig:mag_cryst_struct}
\end{figure}

The pronounced tetragonality, typical of the super-tetragonal structure, allows for a giant polarization of 162.1 $\mu$C/cm$^2$ as estimated from the BECs calculated with VASP and using atomic displacements relative to a reference structure with space group $P4/mmm$, i.e., the closest centrosymmetric structure. 
This calculated value is in good agreement with the previously reported value of 179 $\mu$C/cm²~\cite{Uratani2005}, obtained with \ab~ Berry-phase calculations. 

BiCoO$_3$ also undergoes an antiferromagnetic (AFM) phase transition at the N\'eel temperature $T_{N} = 470~K$ \cite{Belik2006}. 
The magnetic ground state of \bco~has a propagation vector $k= (1/2, 1/2, 0)$, which leads to a C-AFM phase described by a magnetic unit cell with 10 atoms, obtained as a $\sqrt{2}\times\sqrt{2}\times1$ replica of the crystal unit cell described earlier (see Fig. \ref{fig:mag_cryst_struct}). 
We performed structural relaxations, followed by self-consistent total energy calculations at the relaxed structure for different magnetic orders, namely A-AFM, C-AFM, G-AFM, with propagation vector $k = ( 0, 0, 1/2)$, $k = ( 1/2, 1/2, 0)$, $k = ( 1/2, 1/2, 1/2)$, and ferromagnetic (FM).
Our results, summarized in Tab.~\ref{tab:mag_phases}, show that the C-AFM is the most energetically favorable magnetic state in agreement with previous experimental measurements \cite{Belik2006}. 
Furthermore, our calculations with SOC give that BiCoO$_3$ has an easy axis along the $z$ axis (the tetragonal axis), which we will assume through the rest of the manuscript.
The magnetic moment of Co atoms is found to be 3.16/\textit{3.24 }$\mu_{\text{B}}$ (3.0/\textit{3.1} $\mu_{\text{B}}$ from spin  and 0.16/\textit{0.14} $\mu_{\text{B}}$ from orbital moment) respectively with VASP and \textit{ABINIT}.

Formally the d$^6$ configuration leads to a magnetic moment of 4$\mu_{\text{B}}$, but the strong hybridization between the Co d and O p states makes the magnetic moment of Cobalt dilute on the apical site. This result is also seen with muffin-tin spheres calculations \cite{Uratani2005}

\begin{table}[t]
\caption{Energies and lattice parameters of the magnetic phases of \bco~calculated with VASP and ABINIT (in italic font), compared to the experimental values (last row). All the reported values and space groups are for magnetic moments oriented along the $z$ direction.} 
\begin{tabular}{lccrrr}
\hline\hline
\begin{tabular}[c]{@{}c@{}}Mag Space\\  group\end{tabular} & \begin{tabular}[c]{@{}c@{}}Magnetic\\  Phase\end{tabular} & \multicolumn{1}{c}{E(meV/f.u.)} & \multicolumn{1}{c}{$a$ (\r{A})}  & \multicolumn{1}{c}{$c$ (\r{A})} & \multicolumn{1}{c}{$c/a$}  \\
\hline
$P_{p}4'mm'$              & C-AFM       & 0      & 7.390       & 9.362        & 1.267   \\
$C_{I}m'm'2$              & G-AFM       & 17     & 7.389       & 9.360        & 1.267   \\
$P_{2c}m'm'2$              & A-AFM       & 162    & 7.432       & 9.408        & 1.266   \\
$P4m'm'$              & FM          & 192    & 7.421                          & 9.471                          & 1.276                      \\
                      &             &        & $\textit{7.350}$               & $\textit{9.400}$               & $\textit{1.279}$           \\
\hline
Exp~\cite{Belik2006}& C-AFM     &       & 7.459       & 9.448        & 1.267   \\
\hline
\end{tabular}
\label{tab:mag_phases}
\end{table}

In Tab.~\ref{tab:mag_phases} we also report the magnetic space groups (in the OG setting) for all magnetic phases, assuming the Co magnetic moment aligned along the $z$ direction.
Among the four magnetic states analyzed here, only the FM state allows a linear ME response, which is diagonal by symmetry, i.e. $\alpha_{xx}=\alpha_{yy}\neq\alpha_{zz}$. 
On the other hand, time-reversal combined with a fractional translation implies that a net linear ME response is forbidden by symmetry in any of the AFM states.
To understand this more precisely at the microscopic level, next we analyze the ME response (or the lack thereof) of \bco~ in both the FM and the C-AFM phases.
\section{Electronic and lattice responses}
\label{sec:elec_mag_latt_resp}

As we have discussed in Sec. \ref{sec:theory_ME}, to probe the ME response, we can combine DFPT and finite difference techniques (i.e. by determining the BEC and DMC, and using Eq.~\eqref{eq:alpha_latt_1}) or an applied magnetic field technique and extract the induced polarization versus the field amplitude.
In the following, we apply both of them to BiCoO$_3$ in its C-AFM and FM magnetic phases and we compare the results.
We start from the calculations of the BECs and DMCs.

\subsection{Born effective charges}

\begin{table}[!hptb]
\centering
\caption{Calculated Born effective charge tensor components of BiCoO$_3$, obtained with both VASP and ABINIT (in italic font).
}
\begin{tabular}{c|c|c|cccccc}
\hline\hline
Atom                        & Wyc.              & \multicolumn{7}{c}{Born Effective Charges (e)}                                    \\ \hline
                            &                   &        & \multicolumn{2}{c}{$\partial/\partial \mathcal{E}_{x}$} & \multicolumn{2}{c}{$\partial/\partial \mathcal{E}_{y}$} & \multicolumn{2}{c}{$\partial/\partial \mathcal{E}_{z}$} \\ 
\multirow{3}{*}{Bi}         & \multirow{3}{*}{1a} & $\partial/\partial \tau_{x}$     &   5.08  & (\textit{5.07})  &     0 &                  &     0  &                   \\
                            &                     & $\partial/\partial \tau_{y}$     &      0  &                  &  5.08 & (\textit{5.07})  &     0  &                   \\
                            &                     & $\partial/\partial \tau_{x}$     &      0  &                  &     0 &                  &   3.81 &  (\textit{3.72})  \\ \hline
\multirow{3}{*}{Co}         & \multirow{3}{*}{1b} & $\partial/\partial \tau_{x}$     &   2.63  & (\textit{2.30})  &     0 &                  &     0  &                   \\
                            &                     & $\partial/\partial \tau_{y}$     &      0  &                  &  2.63 & (\textit{2.30})  &     0  &                   \\
                            &                     & $\partial/\partial \tau_{x}$     &      0  &                  &     0 &                  &   3.60 &  (\textit{3.43})  \\ \hline
\multirow{3}{*}{O$_{\text{api}}$}  & \multirow{3}{*}{1b} & $\partial/\partial \tau_{x}$     &  -2.22  & (\textit{-2.13}) &     0 &                  &     0  &                   \\
                            &                     & $\partial/\partial \tau_{y}$     &      0  &                  & -2.22 & (\textit{-2.13}) &     0  &                   \\
                            &                     & $\partial/\partial \tau_{x}$     &      0  &                  &     0 &                  &  -3.26 &  (\textit{-3.19}) \\ \hline
\multirow{3}{*}{O$_{\text{bas}}$}  & \multirow{3}{*}{2c} & $\partial/\partial \tau_{x}$     &  -2.75  & (\textit{-2.62}) &  0.02 & (\textit{-0.14}) &     0  &                   \\
                            &                     & $\partial/\partial \tau_{y}$     &   0.02  & (\textit{-0.14}) & -2.75 & (\textit{-2.62}) &     0  &                   \\
                            &                     & $\partial/\partial \tau_{x}$     &      0  &                  &     0 &                  &  -2.07 &  (\textit{-1.98}) \\ \hline
\end{tabular}
\label{tab:becs}
\end{table}

Our calculated BECs are reported in Tab.~\ref{tab:becs} for both VASP and ABINIT codes. 
There is a substantial agreement between the two codes with maximum discrepancies of 15\% of differences on few components, which might be due to the different radii of the PAW spheres or other atomic data parameters difference. 
We can also observe that all the BEC are diagonal (only the O$_\text{bas}$ has small off-diagonal terms along the $x$ and $y$ directions), i.e., the charge rearrangement due to  atomic displacements takes place along the direction of the atomic displacement.
Further on, BECs allow us to compute the mode effective charge Eq.~\eqref{eq:zbare} of polar phonons, which enter in Eq.~\eqref{eq:alpha_latt_1}.

In the non-centrosymmetric space group $P4mm$, the infrared (IR) and Raman active modes coincide, i.e., a mode is silent or both Raman and IR active.
The 12 zone-center optical phonon modes of the 5-atoms unit cell are classified by symmetry according to the irreducible representations of $P4mm$ in the following way: 3$\Gamma_{1}$+$\Gamma_{3}$+8$\Gamma_5$.
$\Gamma_1$ ($A_1$) modes are IR active along the $z$ direction and non-degenerate, whereas $\Gamma_5$ ($E$) modes are IR active in the $xy$ plane and two-fold degenerate.
The $\Gamma_3$ ($B_1$) mode is silent.
In Tab.~\ref{tab:mode_becs} we report the frequency and the calculated mode effective charge for the IR active modes. 
The values obtained with VASP and ABINIT are consistent with one another.

\begin{table}[t]
\caption{Frequency and mode effective charges of IR active modes of BiCoO$_3$, as calculated from VASP (regular font) and ABINIT (italic font).}
\begin{tabular}{|cc|c|cccccc|}
\hline\hline
\multicolumn{2}{|c|}{\multirow{2}{*}{$\nu$(cm$^{-1}$)}} & \multirow{2}{*}{Character} & \multicolumn{6}{c|}{Mode effective charge (e)}                                                                                                                                                                                     \\ \cline{4-9} 
\multicolumn{2}{|c|}{}                             &                            & \multicolumn{2}{c|}{$x$}                                                           & \multicolumn{2}{c|}{$y$}                                                           & \multicolumn{2}{c|}{$z$}                                       \\ \hline
\multicolumn{1}{|c|}{83}             &  \textit{65}    & $\Gamma_5$                        & \multicolumn{1}{c|}{4.91}  & \multicolumn{1}{c|}{\textit{4.79}} & \multicolumn{1}{c|}{0}     & \multicolumn{1}{c|}{0}                              & \multicolumn{1}{c|}{0}     & 0                               \\ \hline
\multicolumn{1}{|c|}{83}             &  \textit{65}    & $\Gamma_5$                       & \multicolumn{1}{c|}{0}     & \multicolumn{1}{c|}{0}                              & \multicolumn{1}{c|}{4.91}  & \multicolumn{1}{c|}{\textit{4.79}} & \multicolumn{1}{c|}{0}     & 0                               \\ \hline
\multicolumn{1}{|c|}{218}            &  \textit{218}           & $\Gamma_1$                        & \multicolumn{1}{c|}{0}     & \multicolumn{1}{c|}{0}                              & \multicolumn{1}{c|}{0}     & \multicolumn{1}{c|}{0}                              & \multicolumn{1}{c|}{-2.79}  & \textit{\maxime{-}2.71} \\ \hline
\multicolumn{1}{|c|}{220}            &  \textit{216}   & $\Gamma_5$                       & \multicolumn{1}{c|}{-2.10} & \multicolumn{1}{c|}{\textit{-1.50}} & \multicolumn{1}{c|}{0}     & \multicolumn{1}{c|}{0}                              & \multicolumn{1}{c|}{0}     & 0                               \\ \hline
\multicolumn{1}{|c|}{220}            &  \textit{216}   & $\Gamma_5$                        & \multicolumn{1}{c|}{0}     & \multicolumn{1}{c|}{0}                              & \multicolumn{1}{c|}{-2.10} & \multicolumn{1}{c|}{\textit{-1.50}} & \multicolumn{1}{c|}{0}     & 0                               \\ \hline
\multicolumn{1}{|c|}{402}            &  \textit{404}   & $\Gamma_5$                        & \multicolumn{1}{c|}{-5.61} & \multicolumn{1}{c|}{\textit{-5.41}} & \multicolumn{1}{c|}{0}     & \multicolumn{1}{c|}{0}                              & \multicolumn{1}{c|}{0}     & 0                               \\ \hline
\multicolumn{1}{|c|}{402}            &  \textit{404}   & $\Gamma_5$                        & \multicolumn{1}{c|}{0}     & \multicolumn{1}{c|}{0}                              & \multicolumn{1}{c|}{-5.61} & \multicolumn{1}{c|}{\textit{-5.41}} & \multicolumn{1}{c|}{0}     & 0                               \\ \hline
\multicolumn{1}{|c|}{417}            &  \textit{430}   & $\Gamma_1$                        & \multicolumn{1}{c|}{0}     & \multicolumn{1}{c|}{0}                              & \multicolumn{1}{c|}{0}     & \multicolumn{1}{c|}{0}                              & \multicolumn{1}{c|}{5.84}  & \textit{5.37}  \\ \hline
\multicolumn{1}{|c|}{546}            &  \textit{542}   & $\Gamma_5$                        & \multicolumn{1}{c|}{-1.95}  & \multicolumn{1}{c|}{\textit{-1.70}} & \multicolumn{1}{c|}{0}     & \multicolumn{1}{c|}{0}                              & \multicolumn{1}{c|}{0}     & 0                               \\ \hline
\multicolumn{1}{|c|}{546}            &  \textit{542}   & $\Gamma_5$                       & \multicolumn{1}{c|}{0}     & \multicolumn{1}{c|}{0}                              & \multicolumn{1}{c|}{-1.95}  & \multicolumn{1}{c|}{\textit{-1.70}} & \multicolumn{1}{c|}{0}     & 0                               \\ \hline
\multicolumn{1}{|c|}{672}            &  \textit{699}   & $\Gamma_1$                        & \multicolumn{1}{c|}{0}     & \multicolumn{1}{c|}{0}                              & \multicolumn{1}{c|}{0}     & \multicolumn{1}{c|}{0}                              & \multicolumn{1}{c|}{-4.23} & \textit{-4.39}  \\ \hline
\end{tabular}
\label{tab:mode_becs}
\end{table}


\begin{figure*}[!ht]
    \centering   \includegraphics[width=1\textwidth]{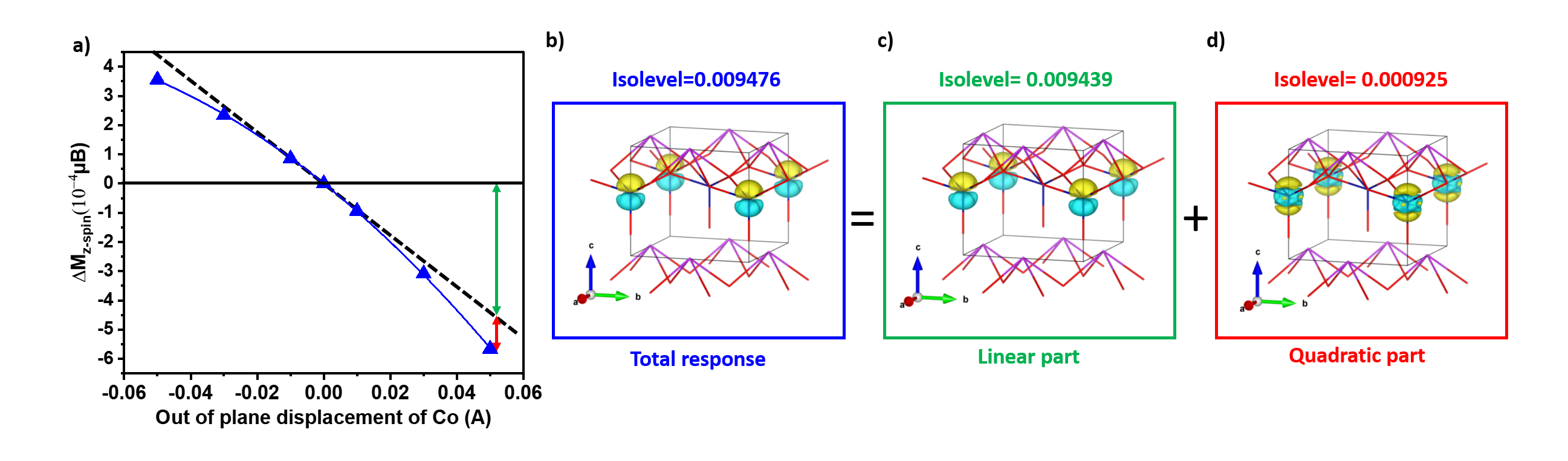}
    \caption{
    (a) Plot of the induced spin magnetization along the $z$ direction versus the amplitude of the Co atom displacement along the $z$ direction. Only the induced magnetization along the $z$ direction is non-zero and it shows a non-linear behavior on top of the linear response (highlighted by the dashed black line. Green and red arrows identify the linear and the non-linear contributions to the induced magnetization, respectively.  Panels b, c, and d, give a visualization of the total, linear and non-linear induced magnetization density, respectively.}
    
    \label{DecompCobaltDMC}
\end{figure*}

\renewcommand{\arraystretch}{1.3}
\begin{table*}[!hptb]
\caption{Linear spin ($\Vec{S}$) and orbital ($\Vec{L}$)  dynamical magnetic effective charges (10$^{-2}\mu_{\text{B}}/\text{\r{A}}$) for each inequivalent Wyckoff sites of \bco\ as calculated with VASP and ABINIT (ABINIT values are in brackets and in italic). 
$\mathcal{H}_i$ refers to the magnetic field in direction $i$ and $\tau_j$ to the atom displacement along the direction $j$.}
\begin{tabular}{c|c|cccccclccccccc}
\hline\hline
    Atom          &     Wyckoff           & \multicolumn{7}{c}{$\Vec{S}$} & \multicolumn{7}{|c}{$\Vec{L}$} \rule{0pt}{1.2em}\\ 

\hline
                  &                   &  & \multicolumn{2}{c}{$\partial/\partial \mathcal{H}_{x}$} & \multicolumn{2}{c}{$\partial/\partial \mathcal{H}_{y}$} & \multicolumn{2}{c|}{$\partial/\partial \mathcal{H}_{z}$} &      & \multicolumn{2}{c}{$\partial/\partial \mathcal{H}_{x}$} & \multicolumn{2}{c}{$\partial/\partial \mathcal{H}_{y}$} & \multicolumn{2}{c}{$\partial/\partial \mathcal{H}_{z}$} \\
\multirow{3}{*}{Bi}   & \multirow{3}{*}{1a} & $\partial/\partial  \tau_{x}$  & 6.93  & (\textit{6.93}) &  0    &       &  0   & \multicolumn{1}{l|}{ } & $\partial/\partial  \tau_{x}$   &  9.83  & (\textit{8.32}) &  0    &       &  0        \\ 
                  &                   & $\partial/\partial  \tau_{y}$  &  0     &            &  -6.93  & (\textit{-6.93})  &  0   & \multicolumn{1}{l|}{} & $\partial/\partial  \tau_{y}$   & 0     &        &      -9.83  & (\textit{-8.32}) & 0  &      \\
                  &                   & $\partial/\partial  \tau_{z}$ &   0     &            &   0     &            & 0 & \multicolumn{1}{l|}{} & $\partial/\partial  \tau_{z}$   &   0     &        &      0  &  & 0  &      \\ \hline
\multirow{3}{*}{Co}   & \multirow{3}{*}{1b} & $\partial/\partial  \tau_{x}$  & 7.25  & (\textit{8.49}) &  0    &       &  0   & \multicolumn{1}{l|}{ } & $\partial/\partial  \tau_{x}$   &  24.63 & (\textit{22.57}) &  0    &       &  0        \\ 
                  &                   & $\partial/\partial  \tau_{y}$  &  0     &            &  7.25  & (\textit{8.49})  &  0   & \multicolumn{1}{l|}{} & $\partial/\partial  \tau_{y}$   & 0     &        &      24.63  & (\textit{22.57}) & 0  &      \\
                  &                   & $\partial/\partial  \tau_{z}$ &   0     &            &   0     &            & -0.91 & \multicolumn{1}{l|}{(\textit{-0.67})} & $\partial/\partial  \tau_{z}$   &   0     &        &      0  &  & -15.01  &  (\textit{-15.34})    \\ \hline
\multirow{3}{*}{O$_{\text{api}}$}   & \multirow{3}{*}{1b} & $\partial/\partial  \tau_{x}$  & 2.80  & (\textit{1.74}) &  0    &       &  0   & \multicolumn{1}{l|}{ } & $\partial/\partial  \tau_{x}$   &  -6.12  & (\textit{-5.52}) &  0    &       &  0        \\ 
                  &                   & $\partial/\partial  \tau_{y}$  &  0     &            &  2.80  & (\textit{1.74})  &  0   & \multicolumn{1}{l|}{} & $\partial/\partial  \tau_{y}$   & 0     &        &      -6.12  & (\textit{-5.52}) & 0  &      \\
                  &                   & $\partial/\partial  \tau_{z}$ &   0     &            &   0     &            & 0.11 & \multicolumn{1}{l|}{(\textit{0.10})} & $\partial/\partial  \tau_{z}$   &   0     &        &      0  &  & -0.57  &   (\textit{-1.83})  \\ \hline
\multirow{3}{*}{O$_{\text{bas}}$}   & \multirow{3}{*}{2c} & $\partial/\partial  \tau_{x}$  & 0 &   &  0    &       &  1.1   & \multicolumn{1}{l|}{(\textit{0.4})} & $\partial/\partial  \tau_{x}$   &  0  &   &  0    &       &  13 &   (\textit{6.1})        \\ 
                  &                   & $\partial/\partial  \tau_{y}$  &  0     &            &   0  &   &    -1.1   & \multicolumn{1}{l|}{(\textit{-0.4})} & $\partial/\partial  \tau_{y}$   & 0     &        &  0 & &   -13 &(\textit{-6.1})             \\
                  &                   & $\partial/\partial  \tau_{z}$ &   -5.8     &      (\textit{-5.9})       &   5.8     &  (\textit{5.9})     & 0 & \multicolumn{1}{l|}{} & $\partial/\partial  \tau_{z}$   &   -3.3     &  (\textit{-3.8})    & 3.3    & (\textit{3.8}) & 0  &      \\ \hline
\end{tabular}
\label{tab:linear_sdmc_odmc}
\end{table*}
\renewcommand{\arraystretch}{1.0}

\subsection{Dynamical magnetic effective charges}
\label{dmc_sec}

We calculated the DMC from finite differences by displacing the atoms along each symmetry-inequivalent directions (See Supplementary Material Fig. S1. for the raw data).
The range of amplitudes chosen allows us to capture linear as well as higher-order, non-linear, contributions to the induced magnetization as exemplified in Fig.~\ref{DecompCobaltDMC} (a) for the case of Co atom.

We establish that most of the responses are linear, but a few of them are significantly non-linear. 
This is the case for the change in orbital magnetization along the $z$ direction ($\Delta m^{\text{orb}}_z$) for Co moving in-plane and the in-plane change of orbital magnetization ($\Delta m^{\text{orb}}_x$ and $\Delta m^{\text{orb}}_y$) for 
O$_{\text{bas}}$ moving in the in-plane direction (see Supplementary Material Fig. S1).

\begin{figure*}[t]
    \includegraphics[width=1\linewidth]{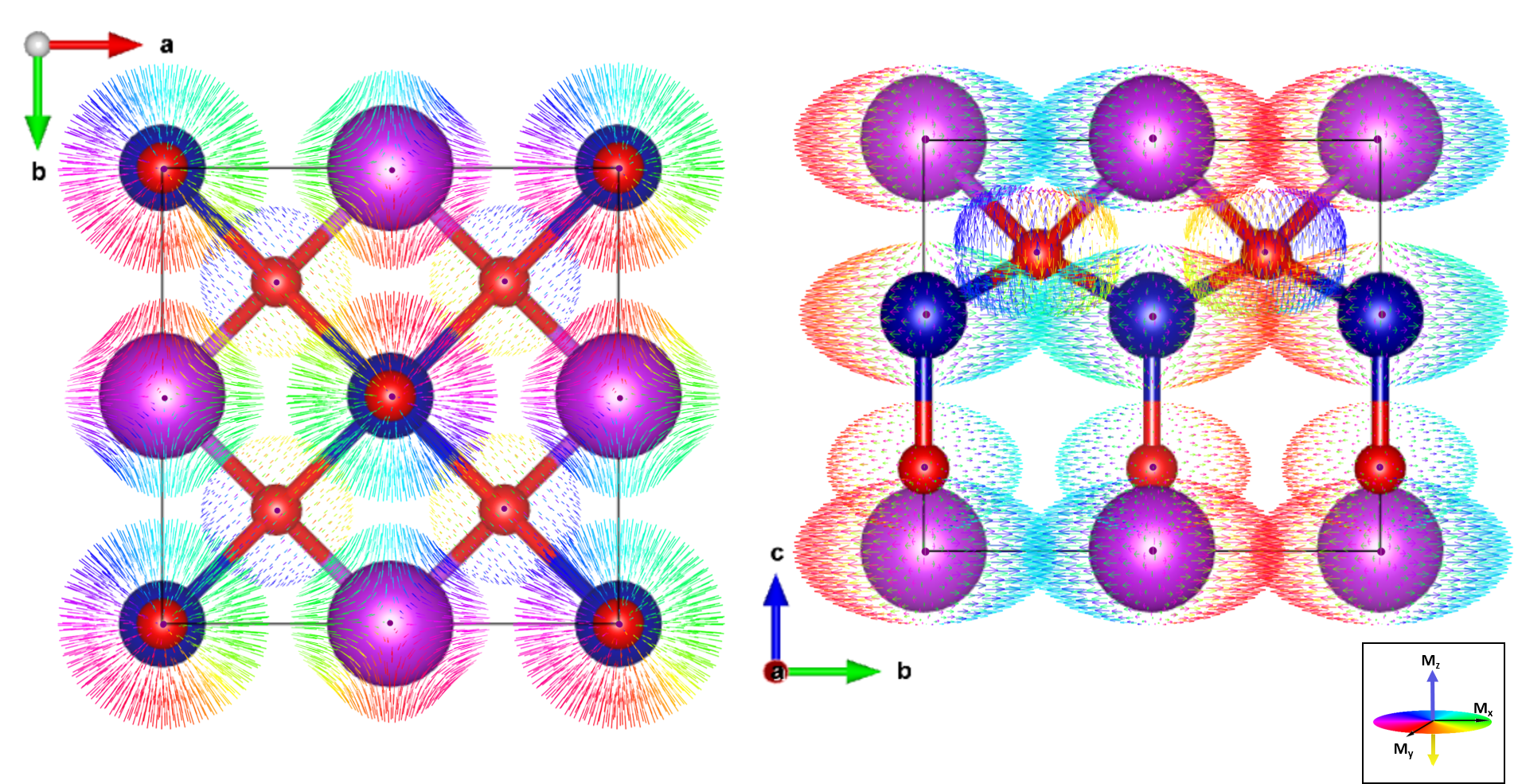}
    \caption{Schematic representation of the DMC pseudotensor as viewed from [001] (left panel) and [100] (right panel) directions. 
    The color of the vectors surrounding Bi, Co and apical Oxygen atom represent the in-plane direction of the induced magnetization when the atom is displaced in a given direction that cross a sphere of arbitrary radius, such that we end up with a 3D vector field for each possible displacement direction around each atom. The vectors surrounding the Basal oxygen are blue when the out-of-plane component of the induced magnetization is along z and yellow for the opposite direction.}
    \label{fig:Magnetizationvectorfield}
\end{figure*}

In Table~\ref{tab:linear_sdmc_odmc} we summarize the spin and orbital contributions to the DMC tensor components, as obtained in both VASP and ABINIT codes from the linear term of the fitted induced magnetization as a function of the atomic displacement. 
Our results show that the orbital contribution is larger than the spin contribution for most of the DMC components. 
This effect is not common and it is particularly enhanced for Co, which can be explained because the Co$^{3+}$ cation with d$^6$ orbital filling is known to potentially have an unquenched orbital moment in crystals. 
Furthermore, we note that the computed DMCs, especially for the orbital contribution, are remarkably larger than those of the paradigmatic ME material, Cr$_2$O$_3$. 
In fact, the largest orbital DMC in BiCoO$_3$ is approximately $0.25 \, \mu_{\text{B}}/\text{\r{A}}$ for Co, two orders of magnitude larger than the largest orbital DMC reported for  Cr$_2$O$_3$ ($0.0025 \, \mu_{\text{B}}/\text{\r{A}}$ for Cr) \cite{Ye2014}.

As explained in Ref.~\cite{Urru2022}, the DMCs are proportional to the local linear ME response. 
For this reason, the shape of the DMC for each atom can be conveniently explained by either (i) the correspondence to the atomic-site ME multipoles introduced earlier in Sec.~\ref{sec:intro}, or similarly (ii) the shape of the local linear ME response, i.e., the linear ME tensor relative to the point symmetry of the Wyckoff site the atom occupies. 
We clarify this point by taking, as an example, the case of the atoms of C-AFM \bco. 
In Table \ref{tab:multipoles} we report the computed ME multipoles for BiCoO$_3$. 
Bi atoms display only a non-zero $x^2-y^2$ ME quadrupole, which results in a diagonal DMC with only one independent parameter, namely $Z^{*\text{m}}_{yy} = -Z^{*\text{m}}_{xx}$ (whereas $Z^{*\text{m}}_{zz} = 0$). 
This corresponds to the shape of the linear ME tensor for Bi's Wyckoff site 2$a$, with local symmetry $4'm m'$. 
Co and the apical O both have the same allowed ME multipoles, namely the ME monopole $a$ and the $z^2$ quadrupole, because they sit in the same Wyckoff site 2$b$. 
Their allowed multipoles, according to Eq.~\eqref{eq_me_tensor}, explain the diagonal shape of the DMC, with two independent parameters $Z^{*\text{m}}_{yy} = Z^{*\text{m}}_{xx}$ and $Z^{*\text{m}}_{zz} \ne 0$, in agreement with the linear ME tensor for the point group $4 m' m'$ (i.e., the site symmetry of Wyckoff position 2$b$). 
Finally, basal oxygens have non-vanishing toroidal moments $t_x$ and $t_y$, as well as $xz$ and $yz$ quadrupoles, which result in the DMC with the shape reported in Table \ref{tab:linear_sdmc_odmc}. 
This has more independent components than the other atoms because the site symmetry of basal O (Wyckoff site 4$c$) is lower ($2'.m'm$).

We remark that DMCs are anisotropic tensors, implying that their shape depends on the orientation of the reference framework.
Practically, if the axes are transformed by a rotation $\mathcal{R}$, the DMCs transform accordingly as $Z^{\prime *\text{m}} = \mathcal{R}^{-1}Z^{*\text{m}}\mathcal{R}$. 

\begin{table}[t]
\caption{Calculated atomic-site ME multipoles (in units of $10^{-2} \, \mu_{\text{B}} \text{\AA}$) for BiCoO$_3$.} 
\begin{tabular}{|c|c|c|c|c|c|c|c|c|c|}
\hline\hline
Atom & Monopole & \multicolumn{3}{|c|}{Toroidal} & \multicolumn{5}{|c|}{Quadrupole} \\
& & \multicolumn{3}{|c|}{moments} & \multicolumn{5}{|c|}{moments} \\
\hline
 & $a$ & $t_x$ & $t_y$ & $t_z$ & $\mathcal{Q}_{x^2-y^2}$ & $\mathcal{Q}_{z^2}$ & $\mathcal{Q}_{xy}$ & $\mathcal{Q}_{xz}$ & $\mathcal{Q}_{yz}$ \\ 
\hline
Bi & -- & -- & -- & -- & $5.8 \times 10^{-3}$ & -- & -- & -- & -- \\
\hline
Co & 1.1 & -- & -- & -- & -- & 1.0 & -- & -- & -- \\
\hline
O$_{\text{api}}$ & - 1.6 & -- & -- & -- & -- & -1.7 & -- & -- & -- \\
\hline
O$_{\text{bas}}$ & -- & 1.8 & 1.8 & -- & -- & -- & -- & 2.3 & -2.3 \\
\hline
\end{tabular}
\label{tab:multipoles}
\end{table}

In Fig.~\ref{fig:Magnetizationvectorfield} we propose a schematic way to visualize the DMCs. 
Particularly, for a given set of points $\textbf{r}$ centered around each atom, with $\textbf{r}$ defined relative to the equilibrium position of the atom, we show with an arrow the magnetization induced by displacing the atom by $\textbf{r}$ and its color is function of its direction. 
Clearly, according to the definition of DMC given in Eq.~\eqref{eq:dmcs}, the induced magnetization is $\Delta M_i = Z^{*\text{m}}_{ij} r_j$, proportional to the DMC, thus the vector fields shown in  Fig.~\ref{fig:Magnetizationvectorfield}  provide a visual representation of the DMCs. 
From Fig.~\ref{fig:Magnetizationvectorfield}, we easily see that Co atom shows a radial and isotropic (similar to an s atomic orbital) magnetization in the $xy$ (i.e., $ab$ in the direct lattice coordinates) plane when shifted from its equilibrium position. 
Furthermore, it shows a magnetization pattern flattened along the $z$ (i.e., $\textbf{c}$) axis. 
Overall, the vector field representing the induced magnetization density has the symmetry of a combined $s$ and $d_{z^2}$ orbital, in agreement with Co showing a spherically symmetric monopole $a$ and a $z^2$ quadrupole, as reported in Table \ref{tab:multipoles}. 
Apical O atoms show a similar behavior to Co atom, to which they form robust bonds characterized by strong hybridization.
This hybridization causes a magnetic moment transfer from  Co atom to O atoms, contributing to the observed magnetization pattern. 
In contrast to Co and apical O, Bi's displacement induces a magnetization with a shape resembling that of a $d_{x^2-y^2}$ atomic orbital, exhibiting lobes with opposite magnetization - one positive and the other negative. 
This is in agreement with the reported $x^2 - y^2$ ME quadrupole on Bi in Table~\ref{tab:multipoles}. 
Finally, basal O induces a chiral/spiral magnetization pattern, justified by its ME toroidal moment.

It is worthwhile noting that the results discussed so far are relative to one magnetic sublattice (the C-AFM contains two magnetic sublattices of opposite direction). 
The DMCs (ME multipoles) for the opposite magnetic sublattice are equal in magnitude but opposite in sign to those reported in Table~\ref{tab:linear_sdmc_odmc} (Table~\ref{tab:multipoles}). 
This happens because the two sublattices are mapped into one another by time-reversal symmetry ($\mathcal{T}$ combined with the $(1/2, 1/2, 0)$ fractional translation is a symmetry of the magnetic lattice), and the DMCs and the ME multipoles change sign under such symmetry. 
On the other hand, the BECs discussed in the previous subsection do not change sign in the two sublattices because they are even under $\mathcal{T}$. 

Having calculated the DMC for each atom, next we use Eq.~\eqref{eq:zbarm} to compute the mode DMCs.
We focus particularly on the phonon modes at zone center, because they are those contributing to the ME response.
First, we note that the 10-atoms C-AFM magnetic unit cell has 30 phonon modes at $\Gamma$, which can be grouped in two sets of 15 modes each. 
The first group contains modes found at the zone center in the 5-atom (i.e., non magnetic or ferromagnetic) unit cell, and thus characterized by in-phase atom displacements for the two magnetic sublattices. 
Among the members of this group there are also IR active, i.e., polar, phonon modes, which show a non-zero mode effective charge as we discussed earlier in Table~\ref{tab:mode_becs}.
The second set, instead, contains modes at the zone-boundary high-symmetry point $M$ ($\textbf{k} = (1/2, 1/2, 0)$) of the 5-atom unit cell, which are folded to zone-center in the BZ of the magnetic $\sqrt{2} \times \sqrt{2} \times 1$ supercell.
Being zone-boundary modes of the non-magnetic cell, they are characterized by atomic displacements with opposing phases in the two magnetic sublattice cells. 
As such, these modes are non-polar and their mode effective charges is null.
As discussed earlier, each atom in a given magnetic sublattice has an opposite DMC to the corresponding atom in the other sublattice. 
Thus, these opposite contributions cancel each other out in phonon modes of the first group, while they add up in phonon modes of the second group, due to the opposite displacements of atoms belonging to the two sublattices, which cancel the inversion of DMC. 
In Table \ref{tab:mode_dmc} we report the mode dynamical magnetic effective charges for the magnetically active modes, i.e., those at the M zone-boundary point of the non-magnetic cell.
In Table \ref{tab:mode_dmc}  we can see that the $M_1$ mode 
has a nonzero mode magnetic effective charge along the $z$ direction and that the $M_5$ modes give nonzero values in the $xy$ plane. 
All the other modes ($M_2$, $M_3$ and $M_4$) are nill. 
Note that, because they have a degeneracy of two, the projection of the pair of degenerate $M_{5}$ modes depends on the choice of the eigenvector direction of the degenerate subspace, so the amplitude of the dynamical magnetic effective charge must be considered as the sum of the pair. 

Overall, the discussion above implies that polar phonon modes are not magnetically active. This is true, in general, for \textit{any} AFM with magnetic symmetry described by a type-IV magnetic space group.

\begin{table}[t]
\caption{Spin, orbital, and total mode DMCs for phonon modes at the M point of the BZ of the non-magnetic unit cell, which fold to $\Gamma$ in the BZ of the magnetic cell. Values obtained with VASP and ABINIT (italic font).}
\begin{tabular}{|c|c|c|c|c|c|c|c|c|c|}
\hline
\multicolumn{2}{|c|}{$\nu$ (cm$^{-1}$)} & \multicolumn{2}{c|}{Character} & \multicolumn{2}{c|}{Spin} & \multicolumn{2}{c|}{Orbital} & \multicolumn{2}{c|}{Total} \\ \hline
92  & 82   & \multicolumn{2}{c|}{M5 (x2)} & 6.8 & \textit{0.4} & 8.6  & \textit{5.3}  & 15.4 & \textit{5.7}  \\ \hline
146 & \textit{129} & \multicolumn{2}{c|}{M5 (x2)} & 6.0 & \textit{3.8} & 13.4 & \textit{5.4}  & 19.4 & \textit{9.2}  \\ \hline
221 & \textit{224} & \multicolumn{2}{c|}{M1} & 0.4 & \textit{1.1} & 11.0 & \textit{21.9} & 11.4 & \textit{23.0} \\ \hline
309 & \textit{308} & \multicolumn{2}{c|}{M5 (x2)} & 1.2 & \textit{11.7} & 28.1 & \textit{31.1} & 32.3 & \textit{42.8} \\ \hline
347 & \textit{349} & \multicolumn{2}{c|}{M5 (x2)} & 2.8 & \textit{7.8} & 7.3  & \textit{11.3} & 10.1 & \textit{19.1} \\ \hline
633 & \textit{623} & \multicolumn{2}{c|}{M1} & 2.9 & \textit{1.5} & 34.6 & \textit{23.4} & 37.5 & \textit{24.9} \\ \hline
654 & \textit{681} & \multicolumn{2}{c|}{M1} & 1.5 & \textit{0.1} & 16.2 & \textit{0.1}  & 17.7 & \textit{0.2}  \\ \hline
\end{tabular}
\label{tab:mode_dmc}
\end{table}

\begin{table*}[t]
\caption{Allowed magnetic octupoles and corresponding allowed entries of the D2MC tensor for BiCoO$_3$. Note: in $Z^{*m(2)}_{ijk}$ symmetry under exchange of $j$ and $k$ is implied, i.e., if $Z^{*m(2)}_{ijk}$ is allowed, $Z^{*m(2)}_{ikj}$ is allowed as well.} 
\begin{tabular}{|c|c|c|}
\hline\hline
Atom & Allowed octupoles & Allowed D2MC entries \\
\hline
Bi & $\mathcal{O}_2$, $\mathcal{Q}^{\tau}_{xy}$ & $Z^{*m(2)}_{xxz} = -Z^{*m(2)}_{yyz}$,  $Z^{*m(2)}_{zxx} = -Z^{*m(2)}_{zyy}$ \\
\hline
Co & $\mathcal{O}_0$, $t^{\tau}_z$ & $Z^{*m(2)}_{xxz} = Z^{*m(2)}_{yyz}$,  $Z^{*m(2)}_{zxx} = Z^{*m(2)}_{zyy}$, $Z^{*m(2)}_{zzz}$ \\
\hline
O$_{\text{api}}$ & $\mathcal{O}_0$, $t^{\tau}_z$ & $Z^{*m(2)}_{xxz} = Z^{*m(2)}_{yyz}$,  $Z^{*m(2)}_{zxx} = Z^{*m(2)}_{zyy}$, $Z^{*m(2)}_{zzz}$  \\
\hline
\multirow{2}{*}{O$_{\text{bas}}$} & $\mathcal{O}_{-3} = \mathcal{O}_3$, $\mathcal{O}_{-1} = -\mathcal{O}_1$, & $Z^{*m(2)}_{xxx} = -Z^{*m(2)}_{yyy}$, $Z^{*m(2)}_{xxy} = -Z^{*m(2)}_{yxy}$, $Z^{*m(2)}_{yxx} = -Z^{*m(2)}_{xyy}$, \\
& $\mathcal{Q}^{\tau}_{yz} = \mathcal{Q}^{\tau}_{xz}$ & $Z^{*m(2)}_{xzz} = Z^{*m(2)}_{yzz}$, $Z^{*m(2)}_{zxz} = -Z^{*m(2)}_{zyz}$ \\
\hline
\end{tabular}
\label{tab:octupoles}
\end{table*}

\subsection{Non-linear Dynamical Magnetic Effective Charges}

Fig.~\ref{DecompCobaltDMC} unambiguously illustrates the feasibility of eliciting non-linear  responses where, for wide enough ranges of displacement amplitudes, the FD approach is able to reliably capture higher-order contributions to the induced magnetization. 
This is particularly true for  cases where a linear induced magnetization is forbidden and the quadratic one is the lowest-order response. 
In most cases, the second-order contribution is much smaller than the linear one; those where the second-order response is non-negligible and necessary to precisely describe the induced magnetization are highlighted with dashed lines in Fig. S1 of the Supplementary Material.

The quadratic induced magnetization can be quantified by the so-called second-order dynamical magnetic effective charge (D2MC henceforth). 
This is a second-order analogue of the DMC given in Eq.~\eqref{eq:dmcs}. 
It has been introduced in Ref.~\cite{Urru2022} and it is defined as:
\begin{equation}
    Z^{*\text{m}(2)}_{ijk} = \Omega \frac{\partial^2 M_{i}}{\partial \tau_{j}\partial \tau_{k}}.
\end{equation}
By construction, the D2MC is a third-rank pseudotensor, symmetric upon exchange of $j$ and $k$. 
Given the definitions of the DMC and the D2MC, the induced magnetization $\boldsymbol{\Delta M}$ as a function of the displacement $\boldsymbol{\tau}$ reads 
\begin{equation}
  \Delta M_i = Z^{*\text{m}}_{ij} \tau_j + Z^{*\text{m}(2)}_{ijk} \tau_j \tau_k, 
  \label{eq:2ndorder}
\end{equation}
thus $Z^{*\text{m}(2)}$ can be obtained from a quadratic fit of $\boldsymbol{\Delta M} (\boldsymbol{\tau})$. 
We note here that while displacements along pure Cartesian directions allow to obtain the full DMC, the same does not apply for the D2MC, and in fact only the entries with $j = k$ can be obtained in this way. 
To obtain the entries with $j \ne k$, linear combinations of displacements along different Cartesian directions are required. 

While a quantitative evaluation of the full D2MC tensors is beyond the scope of the present work, we remark that their allowed entries are determined by symmetry and, similarly to the DMC discussed earlier, can be identified by a link with specific multipoles of the magnetization density. 
As explained in Ref.~\cite{Urru2022}, the D2MCs are linked to the so-called magnetic octupoles, which represent the next-order term in the expansion of Eq.~\eqref{eq:dipole} and are identified by the third-rank pseudotensor $\mathcal{M}_{ijk} = \int d^3 \mathbf{r} \mu_i (\mathbf{r}) r_j r_k$. 
Our computed octupoles in BiCoO$_3$ are reported in Table~\ref{tab:octupoles}, following the naming convention of Ref. \cite{Urru2022}. 
In the same Table, we indicate also the independent entries of the D2MCs implied by the allowed octupoles.

As a final note, we remark that similarly to how the DMCs contribute to the linear ME response (Eq.~\eqref{eq:alpha_latt_1}), the D2MCs contribute to identify the second-order ME response $\beta_{ijk}$, induced by an electric field, introduced in Eq.~\eqref{eq:ME}. 
For a detailed discussion and the derivation we refer the interested reader to the Supplementary Material of Ref.
~\cite{Verbeek2023}. 
Although a quantitative analysis of $\beta_{ijk}$ for each atom of BiCoO$_3$ is not a target of this work, it is important to note that, similarly to the linear ME response, the second-order response is allowed locally, i.e., at the atomic level, but vanishes at the unit cell level, due to the time-reversal symmetry combined with a fractional translation connecting the two sublattices. 
In fact, magnetic octupoles and, similarly, the D2MCs as well as $\beta_{ijk}$, change sign under time-reversal, thus resulting in a second-order anti-ME response, i.e., with equal magnitude but opposite sign in the two sublattices.

\section{Magnetoelectric response}
\label{sec:me_response}

Having calculated the BECs and the DMCs, we can compute the lattice contribution to the ME response using Eq.~\eqref{eq:alpha_latt_1}.
According to Eq.~\eqref{eq:alpha_latt_1}, to have a large lattice-mediated  ME response, a material needs to exhibit low-lying IR phonon modes with correspondingly large mode effective charges and large dynamical magnetic effective charges.

\subsection{C-AFM phase}

\bco~cast large mode effective electric and magnetic charges but not on matching phonon modes.
Indeed, as reported above, the mode DMC  is zero for the polar modes, hence the total ME response is strictly zero.
Let us look more closely at why the ME response vanishes.

\begin{figure}
    \includegraphics[width=0.95\linewidth]{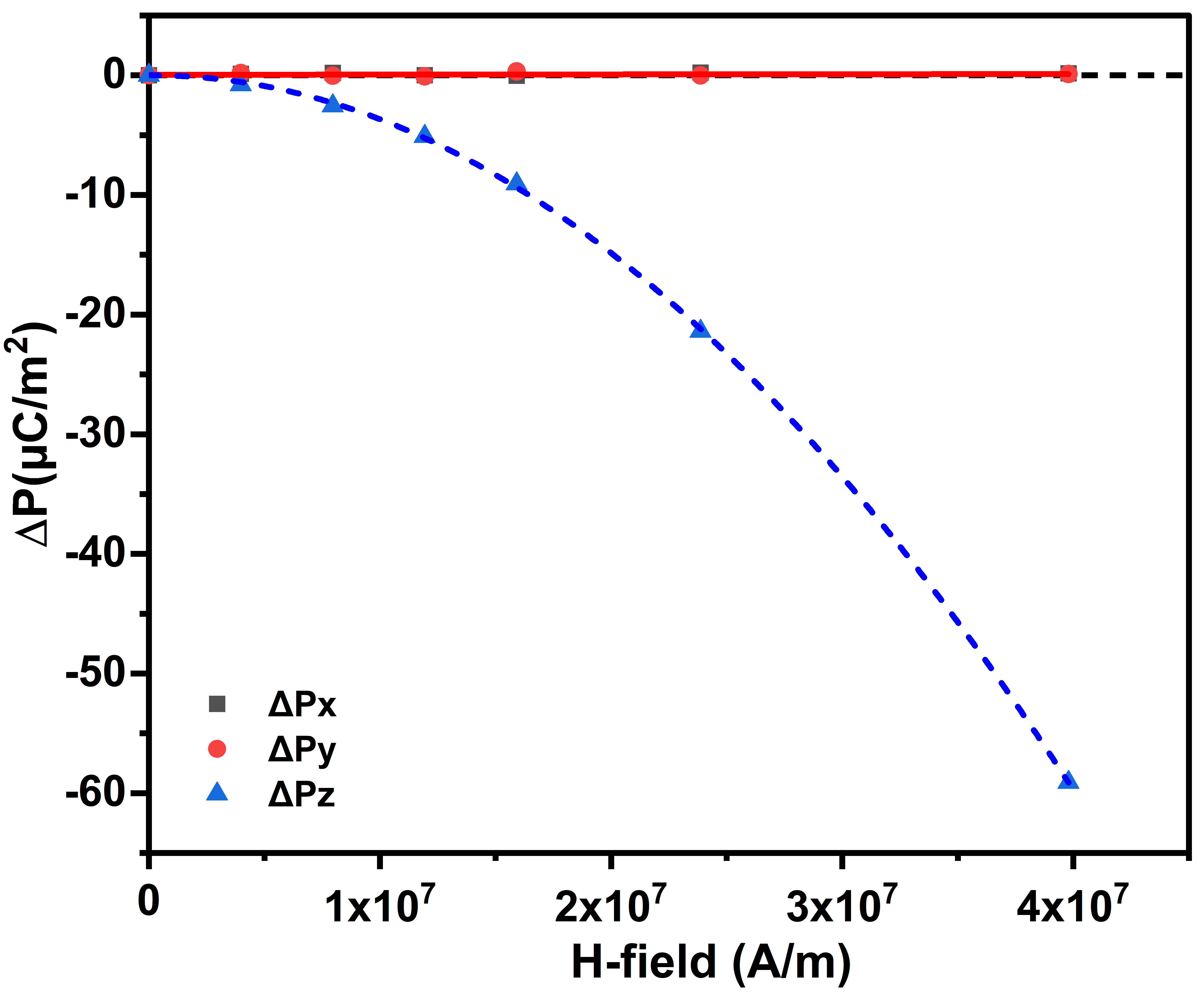}
    \caption{Change of polarization (along the three directions, i.e. $\Delta P_x$, $\Delta P_y$ and $\Delta P_z$) versus applied magnetic field amplitude along the $x$ direction as calculated in the C-AFM phase of BiCoO$_3$ with ABINIT. Only the $\Delta P_z$ gives a response that is non-linear only.}
    \label{fig:CAFMMagField}
\end{figure}

At first sight it can be surprising that \bco\ hosts large atomic DMCs, which result nonetheless in a null total ME response. 
This can be understood by decomposing the total response into the response of each magnetic sublattice. 
The time-reversal symmetry combined with the fractional translation $(\frac{1}{2}, \frac{1}{2}, 0)$ implies that the C-AFM order is described by a propagation vector $\mathbf{q} = (\frac{1}{2}, \frac{1}{2}, 0)$. 
The resulting phase factor describing the magnetization density and, likewise, the induced magnetization, implies that the DMCs are opposite for the two sublattices, while the BECs have the same sign and magnitude. 
Thus, according to Eq.~\eqref{eq:alpha_latt_1}, the two sublattices have non-vanishing, and opposite in sign ME responses, resulting overall in a net vanishing ME response. This qualifies BiCoO$_3$ as an instance of the recently introduced anti-ME case~\cite{Verbeek2023}. 
We find that the amplitude of the (lattice-mediated) ME response of each sublattice is $\alpha_{xx}=85$ ps/m and  $\alpha_{zz}=1$ ps/m.
The $\alpha_{zz}$ response is nonzero through the orbital contribution only, the spin response being zero as  expected~\cite{mostovoy2010, Scaramucci2012}.
We remark that if the magnetic unit cell was the same as the crystallographic one, the propagation vector would be $\mathbf{q} = (0,0,0)$ and hence the ME response would be nonzero.

In Fig.~\ref{fig:CAFMMagField} we report the change of polarization of the C-AFM phase of BiCoO$_3$ under an applied Zeeman magnetic field along the $x$ direction as obtained with the ABINIT code.
We confirm that the linear ME response is strictly zero as no response is observed along the $x$ and $y$ directions ($\Delta P_x$ and $\Delta P_y$ in Fig.~\ref{fig:CAFMMagField}) and a pure quadratic response is observed along the $z$ direction ($\Delta P_z$, blue curve in Fig.~\ref{fig:CAFMMagField}).
Following Eq.~\eqref{eq:ME}, the quadratic response observed along the $z$ direction corresponds to the $\gamma_{xxz}$ tensor component of the quadratic ME response. 
This should not be confused with the other second-order ME response, represented by the tensor $\beta_{ijk}$ discussed in the previous section. 
In fact, $\beta_{ijk}$ identifies the magnetization induced at second order by an applied electric field, whereas $\gamma_{ijk}$ corresponds to the polarization induced at second order by an applied magnetic field. 
In the C-AFM phase of BiCoO$_3$ $\beta_{ijk}$ is allowed only locally, i.e., separately in each sublattice, but vanishes globally because of the time-reversal symmetry combined with fractional translation. 
On the other hand, $\gamma_{ijk}$ is globally allowed by symmetry because the lattice of BiCoO$_3$ breaks inversion symmetry.

\subsection{Ferromagnetic phase: giant magnetoelectric response}

\begin{figure}[tb]
    \includegraphics[width=1\linewidth]{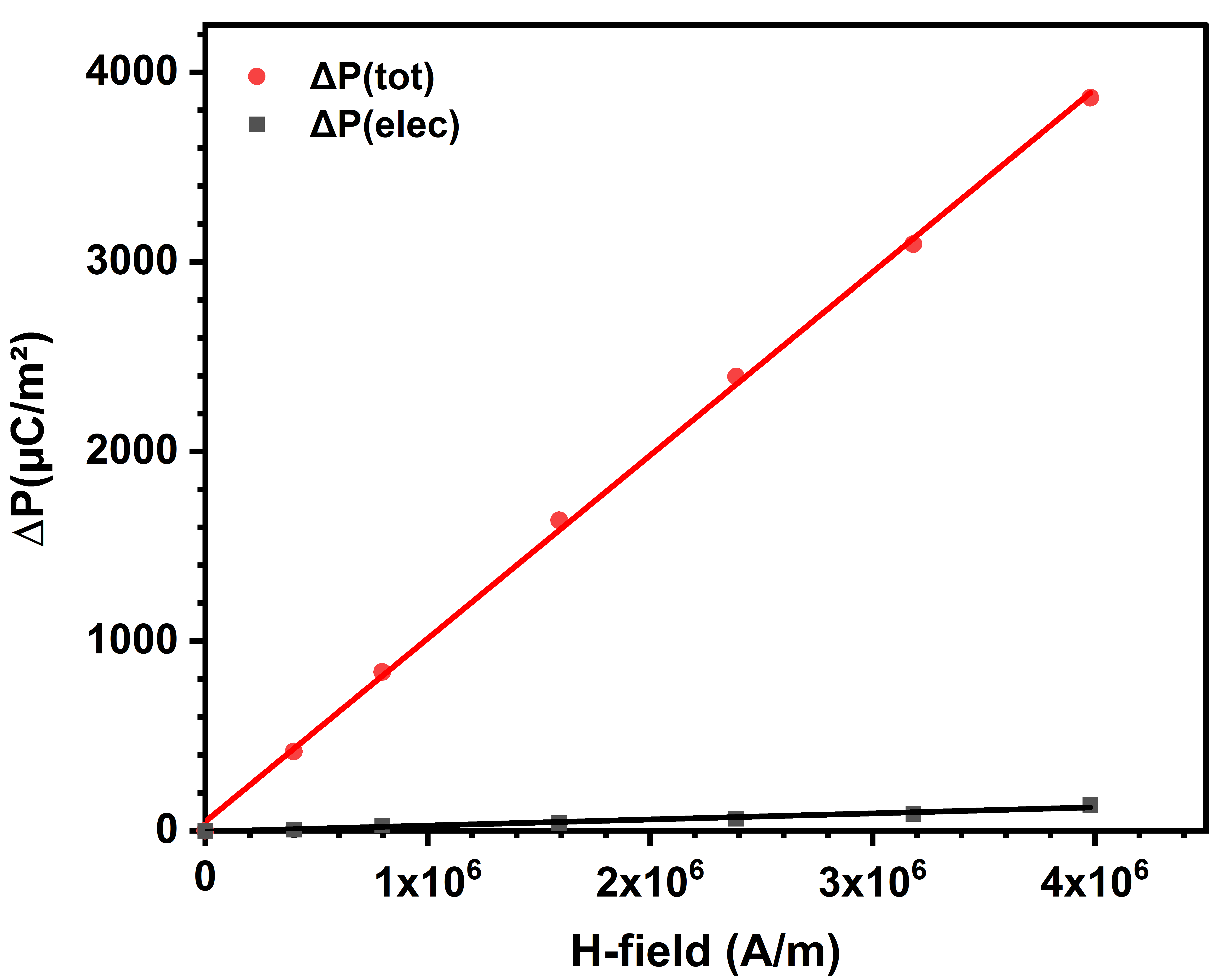}
    \caption{Plot of the change of polarization along the $x$ direction (total in red circles and electronic response only in black squares) of the FM phase of \bco\ under an applied magnetic field along the $x$ direction, hence probing the $\alpha_{xx}$ ME tensor response.}
    \label{fig:FMMagField}
\end{figure}

As we have seen above, the total linear ME response of BiCoO$_3$ is zero because the local responses of each magnetic sublattice perfectly cancel each other due to the C-AFM order.
Here we explore by a computer experiment the ME response of BiCoO$_3$ with an artificial FM order.
The FM order does not lead to any crystal unit-cell doubling because it breaks also the time-reversal symmetry, which (combined with a fractional translation) was instead preserved in the C-AFM phase. 
Hence, we expect a net non-zero ME response in this case.
Assuming spins aligned along the z direction, the FM phase is described by the magnetic space group $P4m'm'$, which allows for a diagonal ME tensor with $\alpha_{xx}=\alpha_{yy}\neq \alpha_{zz}$.

In Fig.~\ref{fig:FMMagField} we report our calculated ME response for the FM phase under an applied Zeeman magnetic field along the $x$ direction (as obtained with the ABINIT code) where the electronic response (clamped ions) and the full response is shown.
The full response is very large and gives a ME response $\alpha_{xx}=1048$ ps/m, i.e., two orders of magnitude larger than what we found for the individual ferromagnetic chains of the C-AFM order. 
Remarkably this response is among the largest bulk (i.e., not composites \cite{vaz2010}) crystal's ME response reported so far in type I multiferroics: among them we mention the ME response of 480 ps/m for Fe$_2$Mo$_3$O$_8$~\cite{Chang2023}, 730 ps/m for TbPO$_4$~\cite{Rivera2009}, approximately 300 ps/m for DyFeO$_3$ (beyond the critical magnetic field of 2.5T at 3 K)~\cite{tokunaga2008}, and approximately 1300 ps/m for Ni$_3$TeO$_6$ (at 2K)~\cite{oh2014}.
The electronic contribution to the response is much smaller, about 16 ps/m, but it is nevertheless one order of magnitude larger than those previously reported in the literature: 0.34 ps/m for Cr$_2$O$_3$, 1.1 ps/m for LiNiPO$_4$~\cite{Bousquet2010}, 1.8 ps/m for FeS~\cite{ricci2016}, and 1.8 ps/m for epitaxially strained CaMnO$_3$~\cite{bousquet2011a}. 

The large electronic response can be attributed to the fact that the band gap is significantly reduced in the FM phase from 1.42 eV in the C-AFM phase to 0.82 eV, giving a large electronic contribution to the permittivity ($\varepsilon_{\infty}$)~\cite{bousquet2011a}.
The large spin contribution to the response can be attributed to the large magnetic susceptibility of the FM phase with respect to the AFM phases (which increases the upper bound for the ME response, Eq.~\eqref{eq:upperbond}). 
Alternatively, it could be explained by the fact that when applying an electric field, the only energy term to counteract in a FM phase is the magnetocrystalline anisotropy energy (MAE), calculated to be approximately 3 meV in our case, whereas in the AFM phase superexchange interactions, which are much larger than the MAE (about 31 meV in the present case), must be kept into account as well.

We cross-checked our result by computing the lattice-mediated ME response using Eq.~\eqref{eq:alpha_latt_1}. 
To this end, we note that looking at the DMCs of the FM phase, we find an enormous in-plane spin DMC value for all atoms, as follows: Bi atom with 64 $\times10^{-2} \mu_{\text{B}}/\text{\r{A}}$, Co atom with 1018 $\times10^{-2} \mu_{\text{B}}/\text{\r{A}}$, apical O with $-439 \times10^{-2} \mu_{\text{B}}/\text{\r{A}}$, and basal O with 112 $\times10^{-2} \mu_{\text{B}}/\text{\r{A}}$. 
The orbital contribution to the DMC, although smaller, is significant for Co with 70 $\times10^{-2} \mu_{\text{B}}/\text{\r{A}}$ and apical O with $-24 \times10^{-2} \mu_{\text{B}}/\text{\r{A}}$.
Altogether, using the finite difference technique, we obtain a lattice contribution to the ME response $\alpha_{xx}$ of 1115 ps/m, with a spin contribution of 1104 ps/m, which is in good agreement with our applied magnetic field calculations.
The out-of-plane component, $\alpha_{zz}$, is found to be non-zero but considerably smaller, around $-7$ ps/m.

The ground state of \bco\ is not FM, but, if one would be able to stabilize BiCoO$_3$ into a FM phase, it could potentially reveal a very large ME response.
In the conclusion, we discuss some strategies to do so.

\section{Conclusions}
\label{sec:conclusions}

We have studied the ME properties and the DMCs of bulk \bco\ in its $P4mm$ supertetragonal ground state structure.
We found that the DMCs have both spin and orbital origins, with the orbital contribution larger than the spin one in the magnetic ground state; hence the orbital contribution cannot be neglected.
We also found that these DMCs are much larger than, e.g., the reference Cr$_2$O$_3$ ME crystal.
We, however, established that BiCoO$_3$ is not ME but rather anti-ME as the ME responses of the two spin sublattices of the C-AFM ground state phase are non-zero but opposite, resulting in a zero total ME response.
We calculated that each ME sublattice carries a rather large ME response amplitude of about 85 ps/m and, as expected from the DMCs, the orbital contribution is significantly larger (roughly 4 times larger) when compared to the spin one.

Because each ferromagnetic sublattice of the C-AFM carries a large ME response, we have computed this response in the higher energy FM phase of BiCoO$_3$ and found a giant response, of 1115 ps/m.
 We argue that the FM phase has a much larger ME response than the single ME sublattice in the C-AFM phase based on the fact that in the FM phase the applied magnetic field needs to overcome the small MAE, whereas in the C-AFM phase an extra barrier is added by the large AFM superexchange.

Our work indicates that by reaching the FM state in \bco\ a very large ME response is unveiled. 
Considering this, let us discuss some of these possibilities.
One possibility is to uncompensate the C-AFM order through doping, i.e., inducing ferrimagnetism, or to advance the material to a double perovskite with two different B-sites. 
However, the superexchange barrier to be overcome will remain the same in the ferrimagnetic system, although a large ME response would still be possible, as shown by the response of each C-AFM magnetic sublattice.
Another problem would be to preserve the polar distortions and the large magnetic exchange giving rise to room temperature magnetoelectricity~\cite{Kim2019}. 
At last, doping may weaken the strong in-plane coupling between Co atoms that occurs in the C-AFM phase, thus leading to a decrease of $T_{\text{N}}$. 

Another possibility is to grow \bco~into superlattices or thin films~\cite{Li2022}, aiming at making the FM phase energetically favorable, as in, for example, LaMnO$_3$~\cite{gilbert2012,hou2014,gilbert2015,an2017,niu2018,zhong2018,guan2019,liu2019,yao2021,yang2021,lu2022} or LaCoO$_3$~\cite{Fuchs2008}. 
A good candidate for superlattices would be the Bi-based FM crystal BiMnO$_3$~\cite{choi2017}.
As a possible route to switch to a FM order, we also mention the anionic substitution, as it was done for, e.g., NaMnF$_3$~\cite{pren2021}, La$_{0,7}$Ca$_{0,3}$MnO$_{3-\delta}$F$_{x}$~\cite{ALTINTAS2014}, and SrNbO$_{3-x}$N${_x}$~\cite{Garcia‐Castro2021}.
We also stress that ferromagnetism can be produced at the domain wall between AFM states~\cite{Hirose2017}, hence locally unveiling a large ME response. 
Overall, we hope that our findings and suggestions motivate further investigation of the possible large ME effect in \bco\ and, possibly, similar compounds. 
To this end, we mention that a similar type of response could be present in the twin system PbVO$_3$~\cite{ga2024, bouvier2023, shpanchenko2004} as it shares the same supertetragonal ferroelectric structure as \bco\ and the same magnetic order, although with different chemistry and orbital filling of the transition metal cation.

In this article, we have also provided an easy way for a 3D visualization of the DMCs and to identify their multipolar nature. 
This is done by plotting, as a vector field, the magnetic moment induced by displacing each atom separately.
The shape of the vector field provides useful insights on the multipolar character of the DMCs. 
In this way we have been able to find that the Bi DMC has a $x^2-y^2$ quadrupolar character, the Co and apical O DMCs have both a monopolar and a $z^2$ quadrupolar origin, and the basal O DMC has a toroidal and $xz$ and $yz$ quadrupolar nature. 
We confirmed these results, which could be guessed from the aforementioned visual representation of the DMCs, by directly computing from first principles the multipoles of the atomic magnetization density. 

Finally, we also emphasize that, in spite of the complexity of the calculations (non-collinear magnetism within DFT+U with spin-orbit coupling, which gives rise to multiple energy minima~\cite{Bousquet2010, payne2018, payne2019, ponet2023}, and the presence of Co$^{3+}$ cation in its high-spin state) and the small quantities at play (the calculations of the DMCs from finite differences demand to probe magnetization changes ranging from 10$^{-5}$ to 10$^{-2}$ $\mu_{\text{B}}$), we obtain very consistent qualitative and quantitative results with two different codes (VASP and ABINIT), i.e., with different implementations and different pseudopotentials, in line with what was reported recently for the simpler case of Cr$_2$O$_3$~\cite{bousquet2024}.
Remarkably, such an agreement is not initially guaranteed by using the codes with a standard computational setup, i.e., with their usual default parameters.
Indeed, we obtain good convergence and agreements only if the calculations are done with high care concerning not only the numerical convergence parameters (plane wave cut-off, k-point grid for the reciprocal space), but particularly the convergence of the wave functions during the self-consistent cycle due to slow convergence of the magnetization.
Our work, therefore, shows that DFT codes are nowadays reliable enough and precise enough to tackle such very subtle materials properties.

\section*{Acknowledgements}

We thank X. Gonze for fruitful discussions. E.B. and B.G. acknowledges the FNRS and the Excellence of Science program (EOS ``ShapeME'', No. 40007525) funded by the FWO and F.R.S.-FNRS.
EB and MB acknowledges the FNRS CDR project ``MULAN'' No. (J.0020.20).
Computational resources have been provided by the Consortium des \'Equipements de Calcul Intensif (C\'ECI), funded by the Fonds de la Recherche Scientifique (F.R.S.-FNRS) under Grant No. 2.5020.11 and the Tier-1 Lucia supercomputer of the Walloon Region, infrastructure funded by the Walloon Region under the grant agreement n°1910247 and by the High Performance Computing Mesocenter of the University of Lille financed by the University, the Hauts-de-France Region, the State, the FEDER and the University's laboratories through a pooling process.

\bibliography{main}

\clearpage

\end{document}


\title{Supplementary Material: Large dynamical magnetic charges and antimagnetoelectricity from spin and orbital origin in  multiferroic \texorpdfstring{BiCoO$_3$}{}}

\author{Maxime Braun}
\affiliation{Physique Th\'eorique des Mat\'eriaux, QMAT, CESAM, Universit\'e de Li\`ege, B-4000 Sart-Tilman, Belgium}
\affiliation{Univ. Lille, CNRS, Centrale Lille, ENSCL, Univ. Artois, UMR 8181-UCCS-Unité de Catalyse et Chimie du Solide, F-59000 Lille, France}
\author{Bogdan Guster}
\affiliation{Physique Th\'eorique des Mat\'eriaux, QMAT, CESAM, Universit\'e de Li\`ege, B-4000 Sart-Tilman, Belgium}

\author{Andrea Urru}
\affiliation{Department of Physics and Astronomy, Rutgers University, Piscataway, New Jersey 08854, USA}

\author{Houria Kabbour}

\affiliation{Univ. Lille, CNRS, Centrale Lille, ENSCL, Univ. Artois, UMR 8181-UCCS-Unité de Catalyse et Chimie du Solide, F-59000 Lille, France}

\author{Eric Bousquet}
\affiliation{Physique Th\'eorique des Mat\'eriaux, QMAT, CESAM, Universit\'e de Li\`ege, B-4000 Sart-Tilman, Belgium}

\maketitle

\section{Different theoretical formulations of lattice dynamics and lattice-mediated magnetoelectric response}

Here we remind the reader that lattice dynamics is usually formulated in two slightly different ways: 
\begin{enumerate}[(i)]
    \item Using the interatomic force constants (IFC) matrix $C_{\kappa \kappa', i j} = \partial^2 E / \partial \tau_{\kappa i} \partial \tau_{\kappa' j}$, where $E$ is the total energy of the system, and $\tau_{\kappa, i}$ identifies the displacement of atom $\kappa$ along direction $i$, the lattice dynamics equation of motion reads 
    \begin{equation}
       \sum_{\kappa, i} C_{\kappa \kappa', i j} u_{n, \kappa, i} = \omega^2_n m_{\kappa'} u_{n, \kappa', j},
        \label{eq:latt_dyn}
    \end{equation}
    where $n$ labels the different modes and $m_{\kappa}$ is the mass of atom $\kappa$.
    Eq. \eqref{eq:latt_dyn} is a \textit{generalized} eigenvalue problem, with eigenvalues $\omega^2_n$ ($\omega_n$ being the phonon frequency of the $n$-th phonon mode) and generalized eigenvectors $u_{n}$. These obey the following generalized orthonormality condition: 
    \begin{equation}
        \sum_{\kappa, i} m_{\kappa} u_{n, \kappa, i} u_{n', \kappa, i} = \delta_{nn'} 
        \label{eq:norm_u}
    \end{equation}
\item Using the dynamical matrix $D_{\kappa \kappa', i j}$, related to the IFC matrix by $D_{\kappa \kappa', i j} = 1/\sqrt{m_{\kappa} m_{\kappa'}} C_{\kappa \kappa', i j}$, the lattice dynamics equation of motion is
\begin{equation}
    \sum_{\kappa, i} D_{\kappa \kappa', i j} w_{n, \kappa, i} = \omega^2_n w_{n, \kappa', j}.
\end{equation}
This is a regular eigenvalue problem, having the same eigenvalues as Eq. \eqref{eq:latt_dyn}, but different eigenvectors. The eigenvectors $w_{n}$ are orthonormalized in this way: 
\begin{equation}
    w_{n, \kappa, i} w_{n', \kappa, i} = \delta_{nn'}, 
\end{equation}
and are related to the generalized eigenvectors of the IFC matrix in the following way: 
\begin{equation}
    u_{n, \kappa, i} = \frac{w_{n, \kappa, i}}{\sqrt{m_{\kappa}}}.
    \label{eq:eigenvectors}
\end{equation}
\end{enumerate}

The lattice-mediated magnetoelectric (ME) response $\alpha^{\text{latt.}}_{ij}$ was formulated independently in two different, but equivalent, ways in Refs. \cite{Iniguez2008, Ye2014}. In Ref. \cite{Iniguez2008} $\alpha^{\text{latt.}}_{ij}$ reads
\begin{equation}
    \alpha^{\text{latt.}}_{ij} = \frac{\mu_0}{\Omega} \sum_{n = 1}^{N_{\text{IR}}} \frac{\bar{Z}^{m}_{n, i} \bar{Z}^{e}_{n, j}}{C_n},
    \label{eq:alpha_latt_1}
\end{equation}
where $C_n$ is the $n$-th eigenvalue of the IFC matrix, and $\bar{Z}^{e}_{n, i}$ and $\bar{Z}^{m}_{n, i}$
are the mode effective charge and mode dynamical magnetization, respectively (see Eqs. (10) and (11) of the main manuscript). These identify respectively the polarization and the magnetization induced by a phonon mode with amplitude 1 \AA. However, the eigenvectors of the IFC matrix are not normalized in a way that their magnitude is 1, i.e., $|u_{n}|^2 \ne 1$ (see Eq. \eqref{eq:norm_u}), thus they have to be renormalized in this way 
\begin{equation}
    \tilde{u}_{n, \kappa, i} = \frac{u_{n, \kappa, i}}{\sqrt{|u_{n}|^2}}
\end{equation}
in order to be able to define $\bar{Z}^{e}_{n, i}$ and $\bar{Z}^{m}_{n, i}$.

In Ref. \cite{Ye2014}, instead the lattice-mediated ME response is formulated as follows:
\begin{equation}
    \alpha^{\text{latt.}}_{ij} = \frac{\mu_0}{\Omega} \sum_{\kappa \kappa'} \sum_{i' j'} Z^{*\text{m}}_{\kappa, i, i'} (C^{-1})_{\kappa \kappa', i' j'} Z^{*\text{e}}_{\kappa', j', j},
    \label{eq:alpha_latt_2}
\end{equation}
where $Z^{e}_{\kappa, i, i'}$ and $Z^{m}_{\kappa', j', j}$ are the Born effective charge and dynamical magnetic charge tensors of atoms $\kappa$ and $\kappa'$, respectively.

Eq. \eqref{eq:alpha_latt_1} uses mode-specific quantities like the mode effective charge and the mode dynamical magnetization, which are helpful to understand how much each infrared-active mode contributes to the ME response. However, it uses the pure eigenvalue of the IFC matrix, which is different from the generalized eigenvalue (i.e., the phonon frequency) and is not directly related to Eq. \eqref{eq:latt_dyn}. On the other hand, Eq. \eqref{eq:alpha_latt_2} uses the inverse of the IFC matrix, which is directly related to Eq. \eqref{eq:alpha_latt_2}, but does not use mode-dependent effective charges, thus an evaluation of the contribution of each infrared-active mode is not available with this approach.

In this work, we use an intermediate formulation, which combines the advantages of both methods. Our approach is a generalization of the formulation for the dielectric permittivity presented in Ref. \cite{Maradudin_book}, Eq. (6.2.100). Based on this Equation, we generalize it to the ME case in the following way (note that Ref. \cite{Maradudin_book} works with the eigenvectors $w_n$ of the dynamical matrix): 
\begin{equation}
    \begin{split}
    \alpha^{\text{latt.}}_{ij} & = \frac{\mu_0}{\Omega} \sum_{n = 1}^{N_{\text{IR}}} \sum_{\kappa \kappa'} \sum_{i' j'} \frac{Z^{*\text{m}}_{\kappa, i i'}}{\sqrt{m_{\kappa}}} \frac{w_{n, \kappa, i'} w_{n, \kappa', j'}}{\omega_n^2} \frac{Z^{*\text{e}}_{\kappa', j' j}}{\sqrt{m_{\kappa'}}} \\
    & = \frac{\mu_0}{\Omega} \sum_{n = 1}^{N_{\text{IR}}} \sum_{\kappa \kappa'} \sum_{i' j'} \frac{\left(Z^{*\text{m}}_{\kappa, i i'} u_{n, \kappa, i'} \right) \left(Z^{*\text{e}}_{\kappa', j' j} u_{n, \kappa', j'} \right)}{\omega_n^2} \\
    & = \frac{\mu_0}{\Omega} \sum_{n = 1}^{N_{\text{IR}}} \frac{S_{n, i j}}{\omega_n^2},
    \end{split}
    \label{eq:maradudin}
\end{equation}
where in the second line we have used Eq. \eqref{eq:eigenvectors} to express the eigenvectors of the dynamical matrix in terms of the eigenvectors of the IFC matrix, and in the last line we have defined the \textit{ME mode-oscillator strength tensor} as (Eq. (7) in the main manuscript)
\begin{equation}
    S_{n,ij}=\left(\sum_{\kappa,i'}Z^{*\text{m}}_{\kappa,ii'}{u}_{n,\kappa,i'}\right)\times\left(\sum_{\kappa,j'}Z^{*\text{e}}_{\kappa,j'j}{u}_{n,\kappa,j'}\right).
\end{equation}

Next, we prove that our method is equivalent to that of Ref. \cite{Ye2014}. We start from Eq. \eqref{eq:latt_dyn} and express the eigenvectors by applying the inverse of the IFC matrix to both sides. We have: 
\begin{equation}
    u_{n, ,\kappa, i} = \omega^2_n C^{-1}_{\kappa \kappa', i j} m_{\kappa'} u_{n, \kappa', j}.
\end{equation}
After inserting it into our formulation of the lattice-mediated ME response, Eq. \eqref{eq:maradudin}, we get: 
\begin{equation}
    \alpha^{\text{latt.}}_{ij} = \frac{\mu_0}{\Omega} \sum_{n = 1}^{N_{\text{IR}}} \sum_{\kappa \kappa' \kappa''} \sum_{i' j' j''} Z^{*\text{m}}_{\kappa, i i'} C^{-1}_{\kappa, \kappa'', i' j''} Z^{*\text{e}}_{\kappa', j' j} m_{\kappa''} u_{n, \kappa'', j''} u_{n, \kappa', j'}.
    \label{eq:proof}
\end{equation}
From Eq. \eqref{eq:norm_u} we define the completeness relation 
\begin{equation}
    \sum_n m_{\kappa''} u_{n, \kappa'', j''} u_{n, \kappa', j'} = \delta_{\kappa' \kappa''} \delta_{j' j''},
\end{equation}
which, when inserted in Eq. \eqref{eq:proof} gives: 
\begin{equation}
     \alpha^{\text{latt.}}_{ij} = \frac{\mu_0}{\Omega} \sum_{\kappa \kappa'} \sum_{i' j'} Z^{*\text{m}}_{\kappa, i, i'} (C^{-1})_{\kappa \kappa', i' j'} Z^{*\text{e}}_{\kappa', j', j},
\end{equation}
i.e., the formulation of Ref. \cite{Ye2014}.

\section{Supplementary figures}

\begin{figure*}[!hptb]
\centering
\includegraphics[width=.8\linewidth]{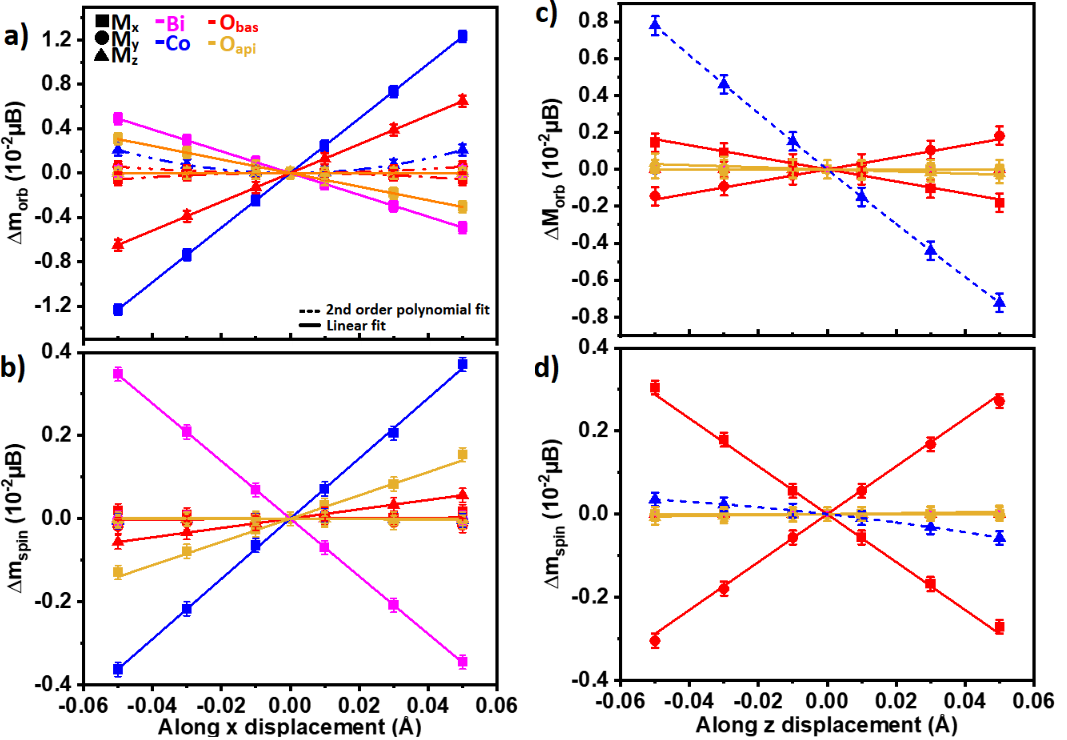}
\caption{Induced orbital (panels a) and c)), and spin (panels b) and d)) magnetization as a function of the amplitude of the displacement of each atom, necessary to compute the DMC using the finite difference technique.}
\label{fig:DMCfit}
\end{figure*}

\begin{figure*}
    \centering
    \includegraphics[width=0.8\textwidth]{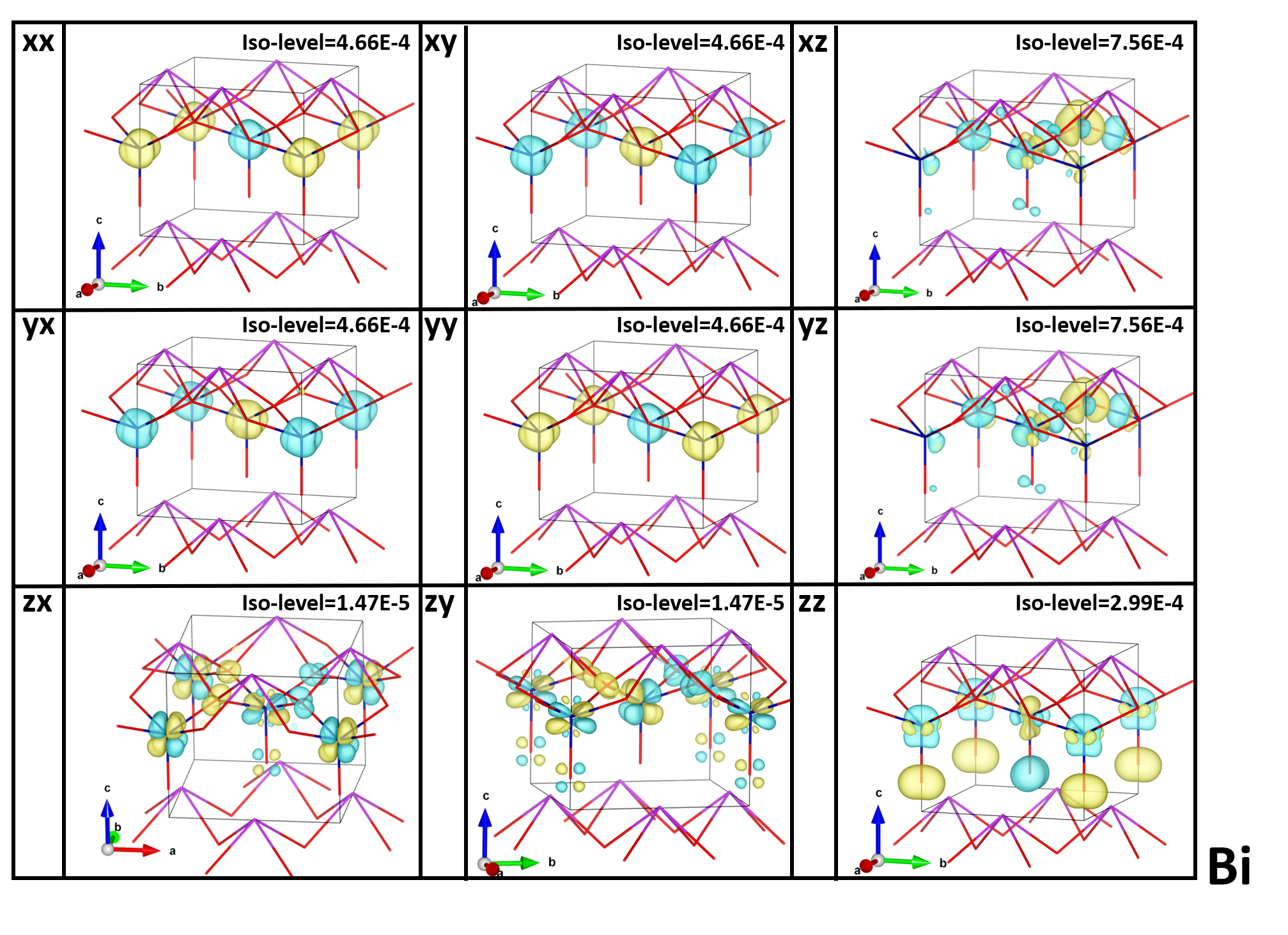}
    \caption{Bismuth DMC Matrix representation}
\end{figure*}

\begin{figure*}
    \centering
    \includegraphics[width=0.8\textwidth]{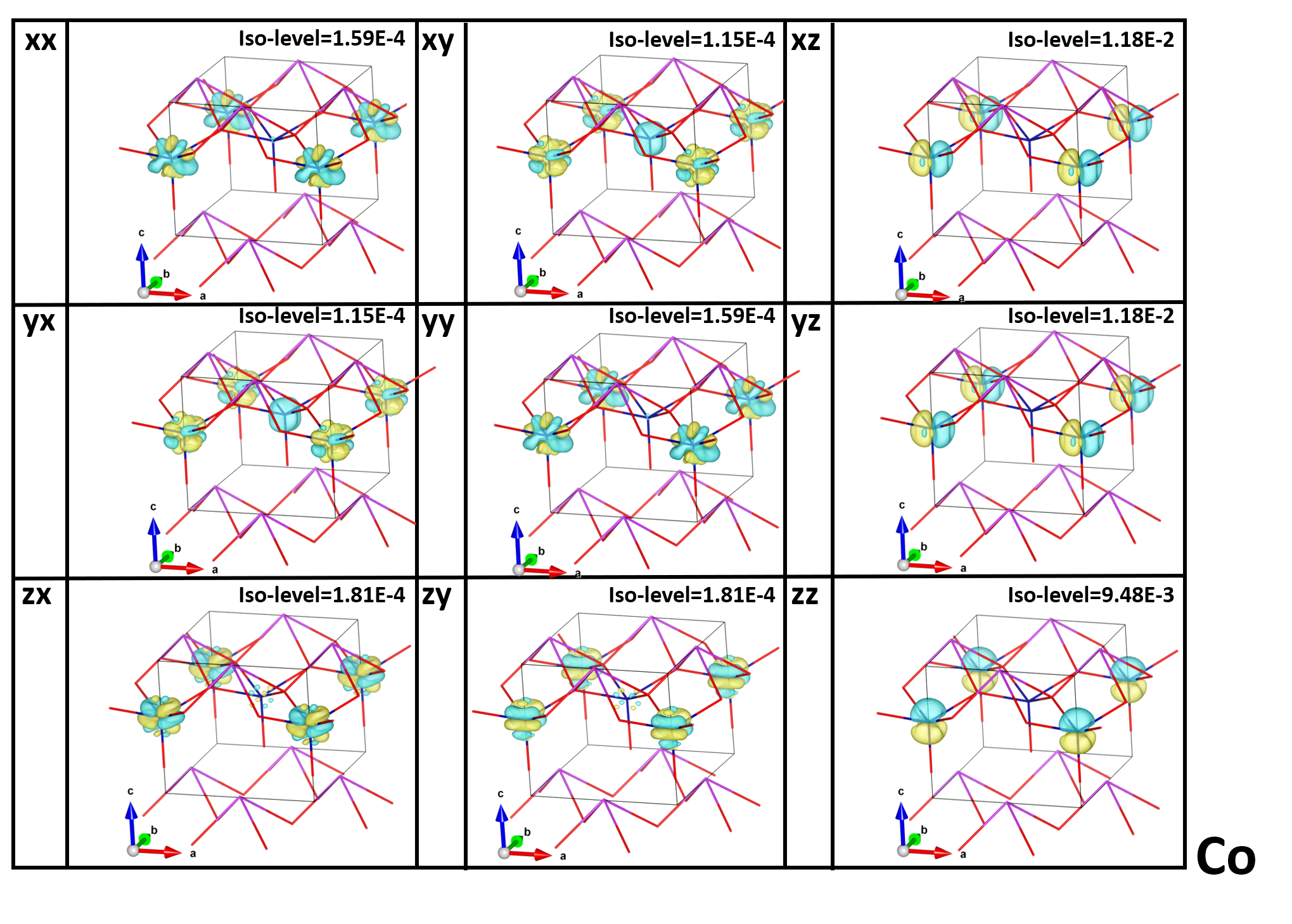}
    \caption{Cobalt DMC Matrix representation}
\end{figure*}

\begin{figure*}
    \centering
    \includegraphics[width=0.8\textwidth]{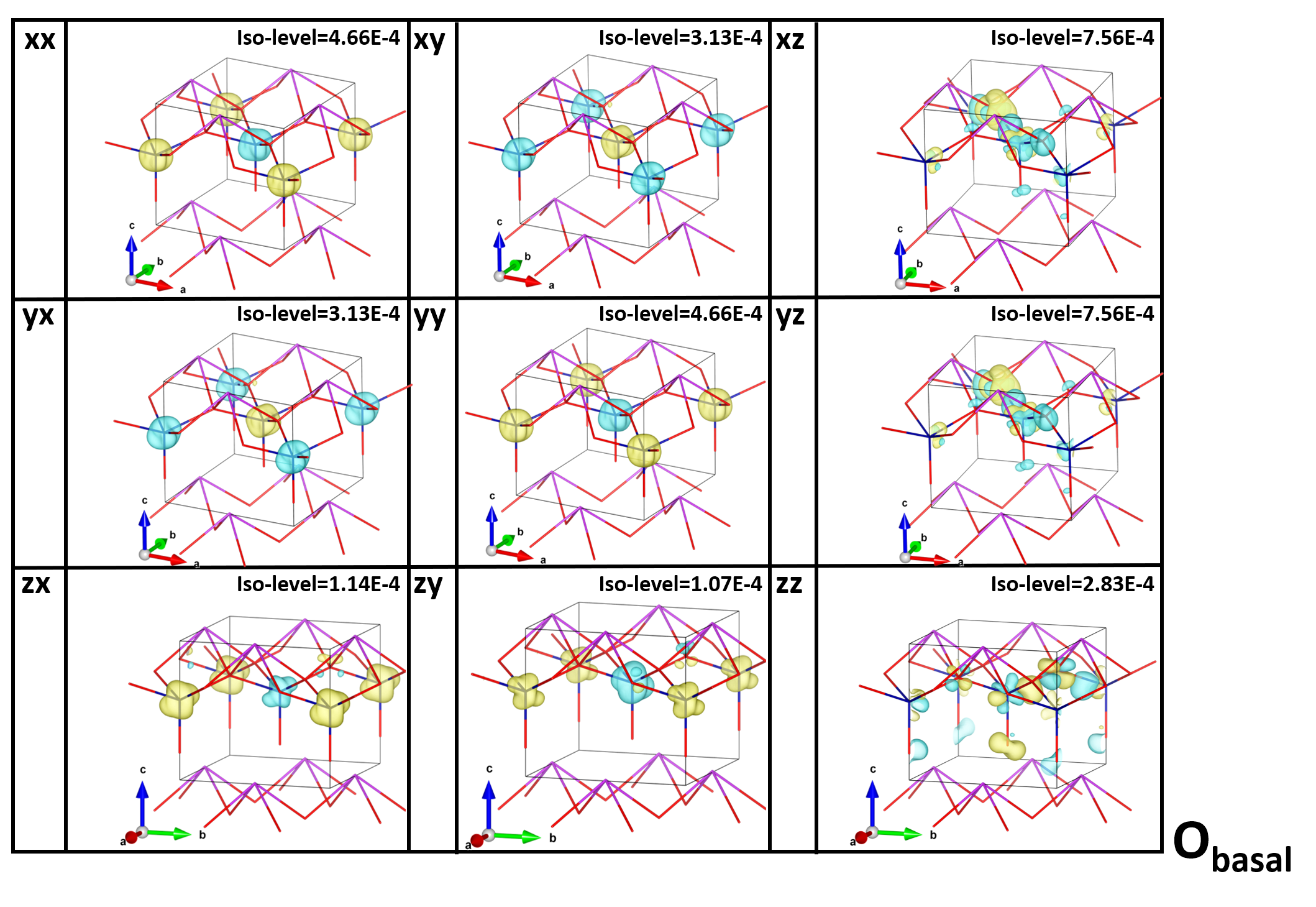}
    \caption{O$_{basal}$ Basal oxygen DMC Matrix representation}
\end{figure*}

\begin{figure*}
    \centering
    \includegraphics[width=0.8\textwidth]{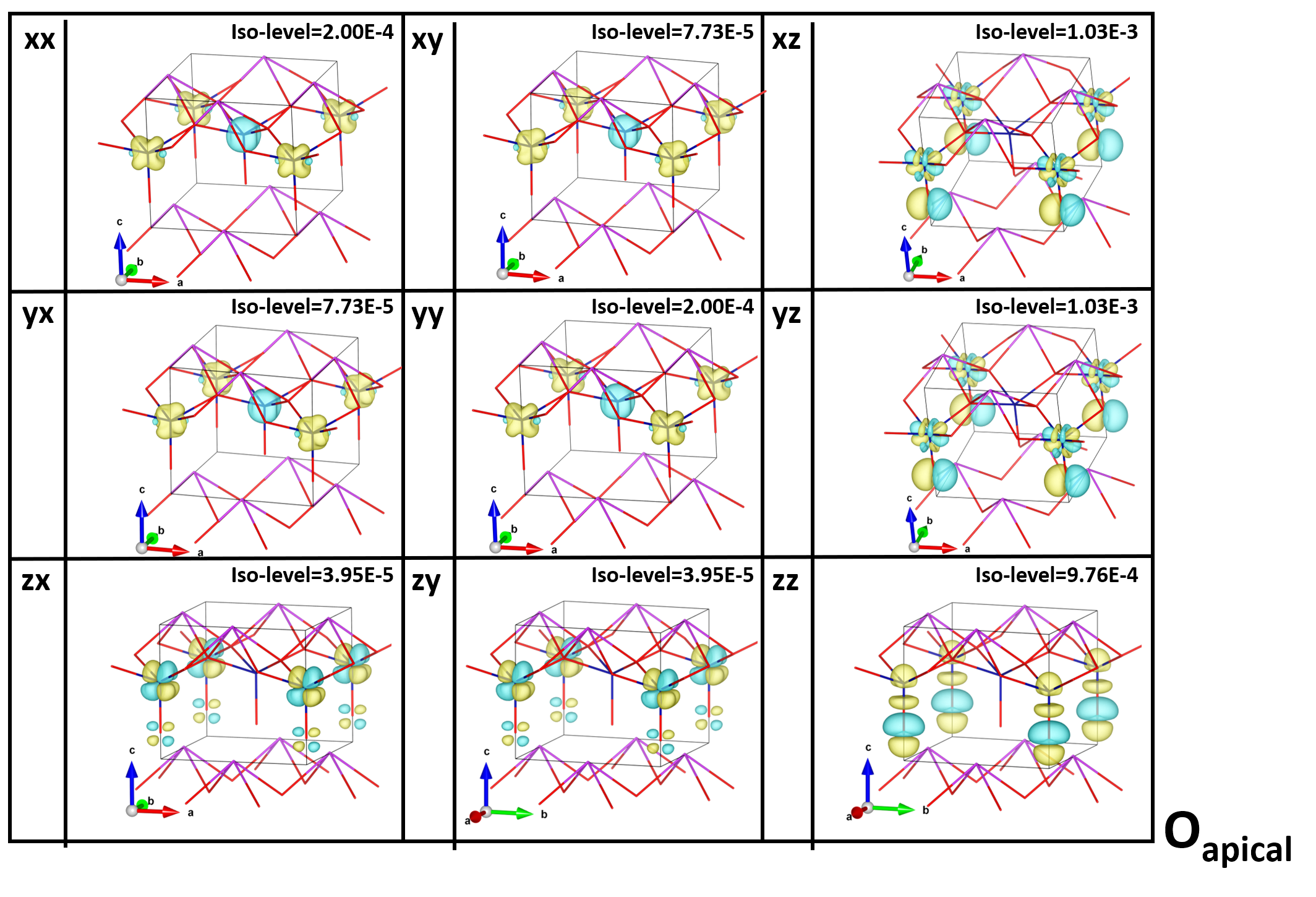}
    \caption{O$_{apical}$ Apical oxygen DMC Matrix representation}
\end{figure*}

\begin{figure*}
    \centering
    \includegraphics[width=0.8\textwidth]{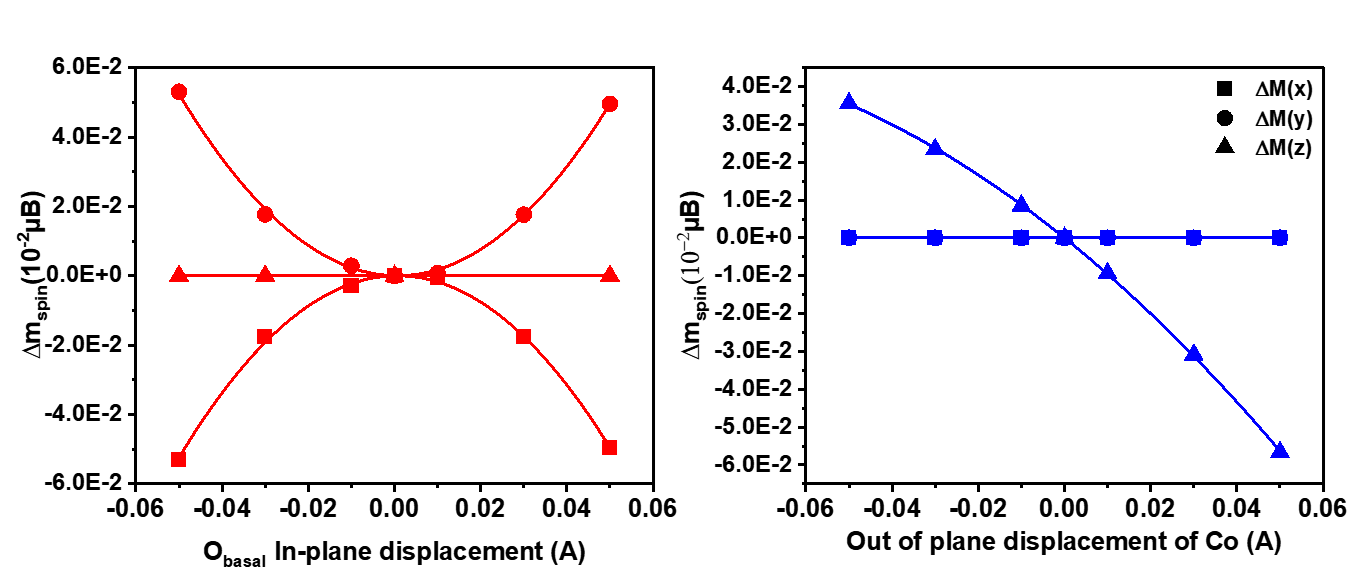}
    \caption{Origin of the second order response in the C-AFM phase}
\end{figure*}

\begin{figure*}
    \centering
    \includegraphics[width=0.6\linewidth]{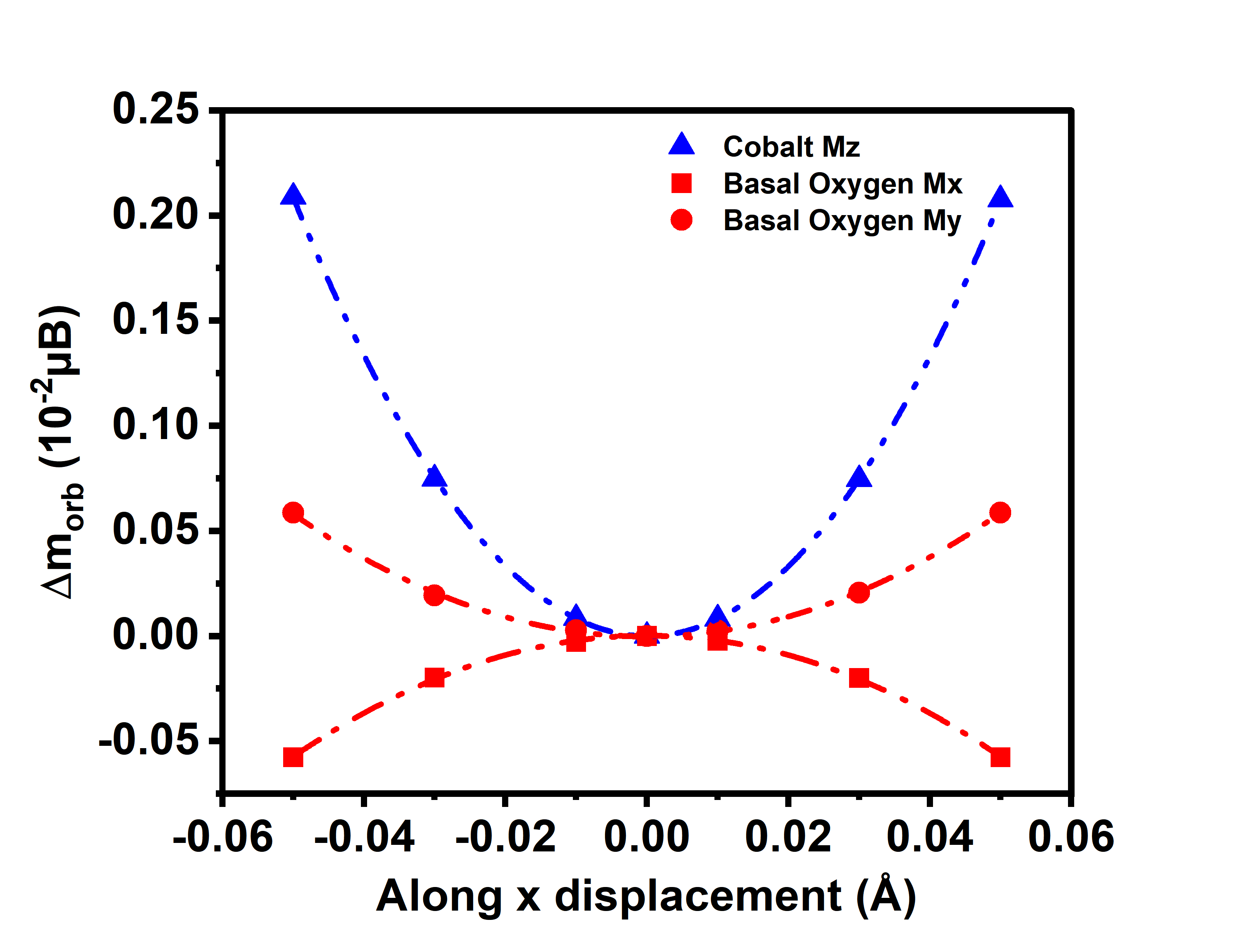}
    \caption{2$^{nd}$ order orbital magnetic response of cobalt and basal oxygen to their displacement in plane}
    \label{fig:NonLinearResponseDMCORB}
\end{figure*}

\begin{figure*}
    \includegraphics[width=0.6\linewidth]{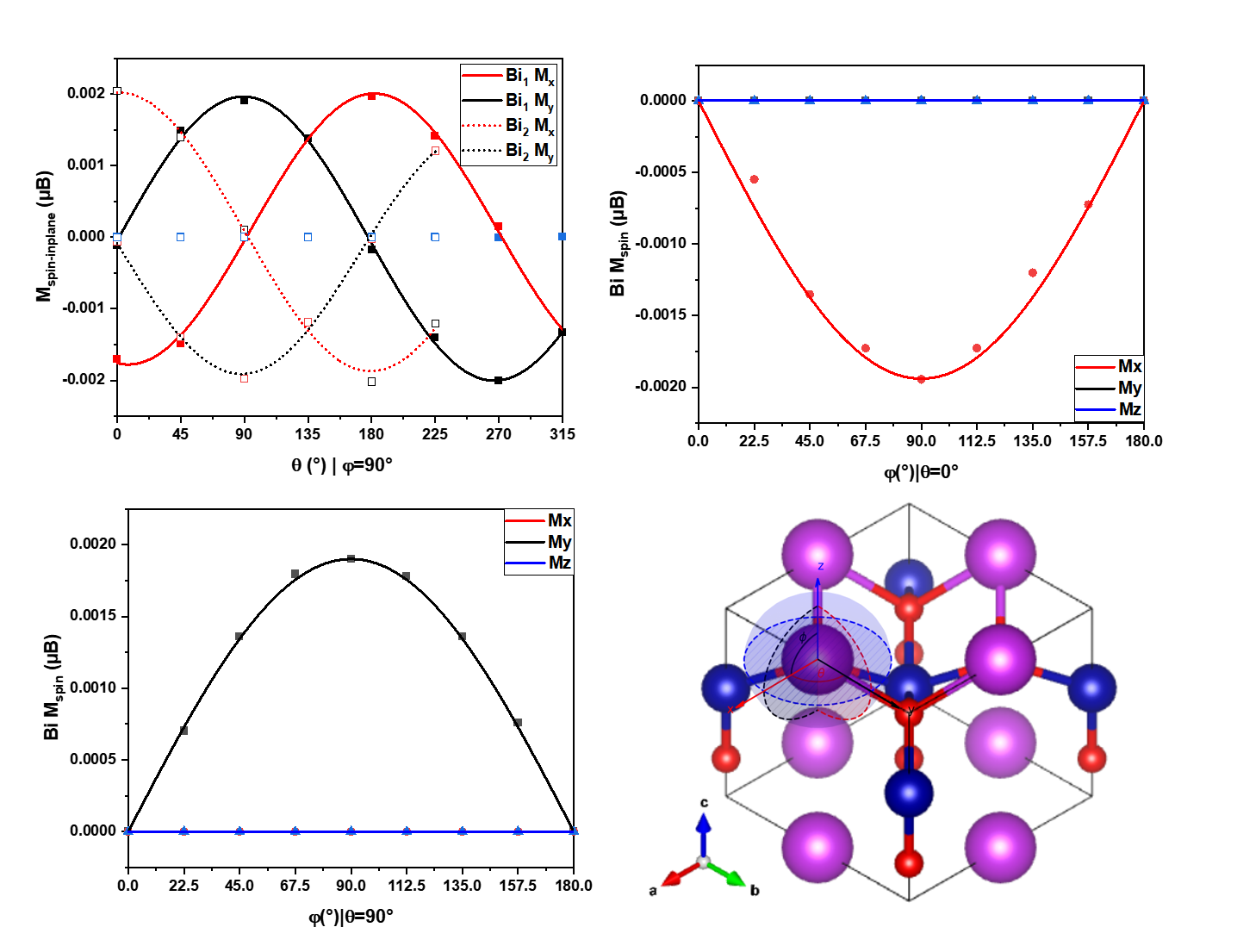}
    \caption{Behavior of the spin magnetization induced by a displacement of the Bi atom, with fixed amplitude ($\tau$=0.03\r{A}), as a function of the direction of the displacement. The direction of the displacement is identified by polar coordinates $\theta$, $\phi$, as shown in the bottom right panel.}
    \label{fig:BismuthVar}
\end{figure*}
\begin{figure*}
    \includegraphics[width=0.6\linewidth]{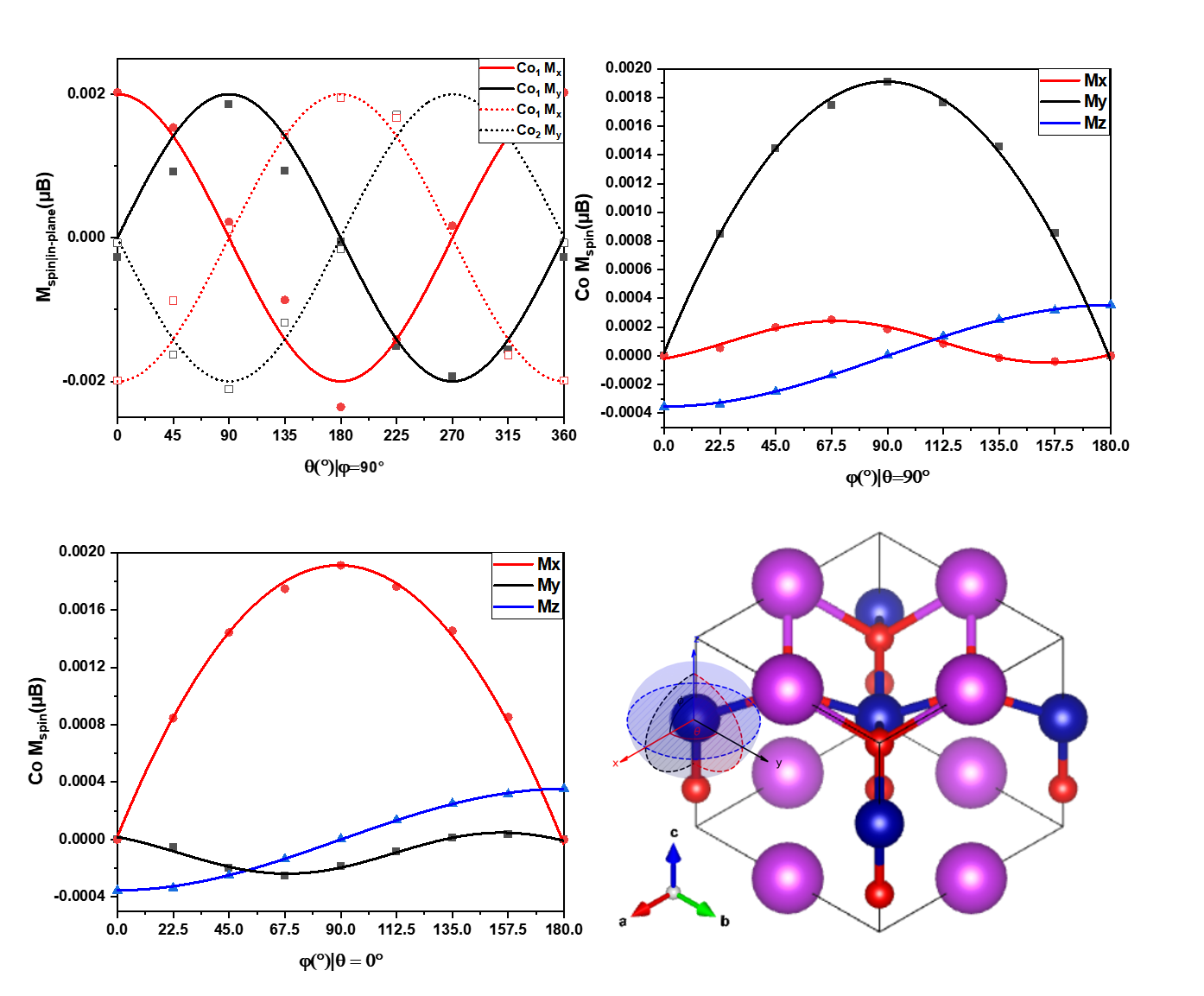}
    \caption{Behavior of the spin magnetization induced by a displacement of the Co atom, with fixed amplitude ($\tau$=0.03\r{A}), as a function of the direction of the displacement. The direction of the displacement is identified by polar coordinates $\theta$, $\phi$, as shown in the bottom right panel.}
    \label{fig:CobaltVar}
\end{figure*}
\begin{figure*}
    \includegraphics[width=0.6\linewidth]{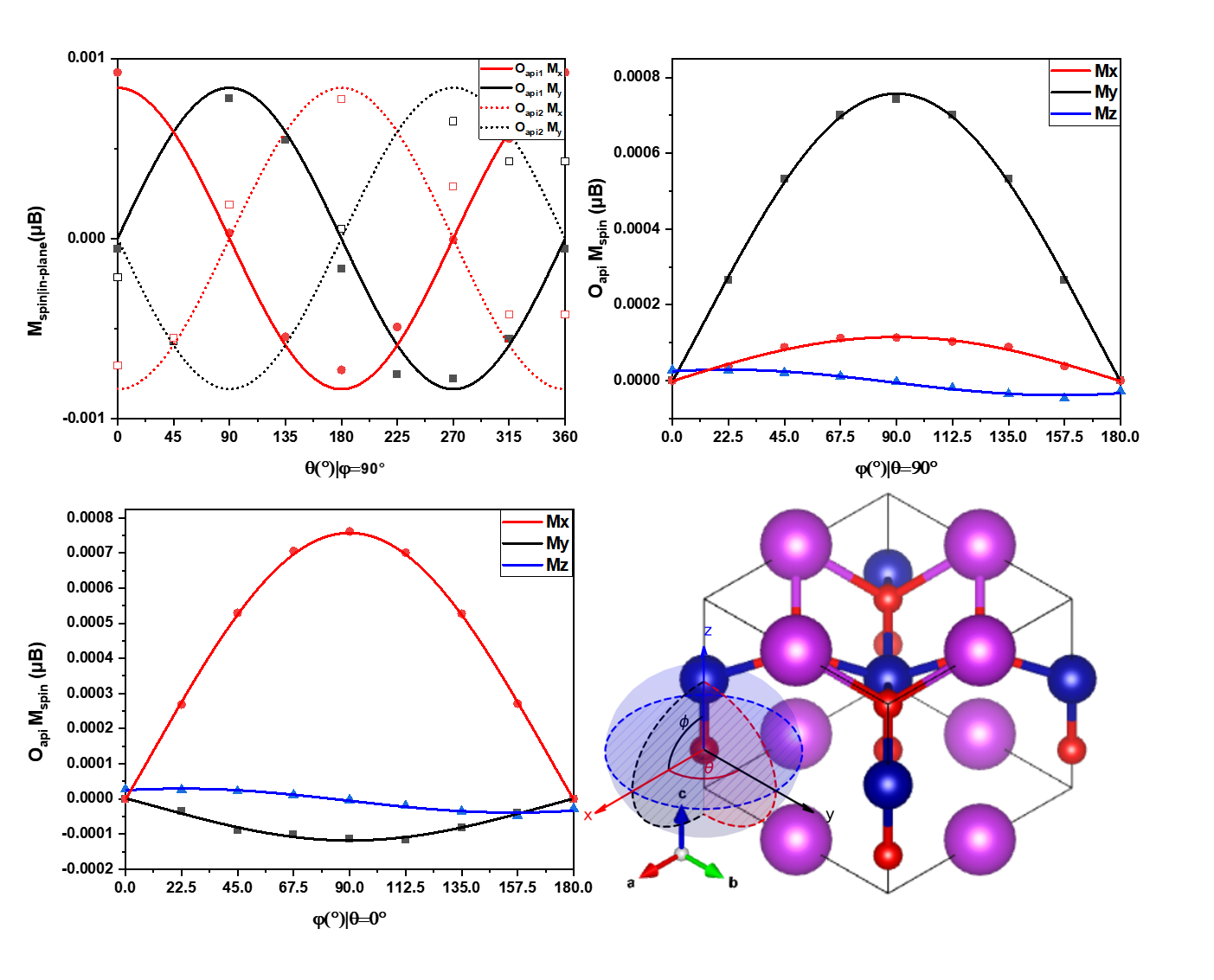}
    \caption{Behavior of the spin magnetization induced by a displacement of the apical oxygen atom, with fixed amplitude ($\tau$=0.03\r{A}), as a function of the direction of the displacement. The direction of the displacement is identified by polar coordinates $\theta$, $\phi$, as shown in the bottom right panel.}
    \label{fig:OapiVar}
\end{figure*}
\begin{figure*}
    \includegraphics[width=0.6\linewidth]{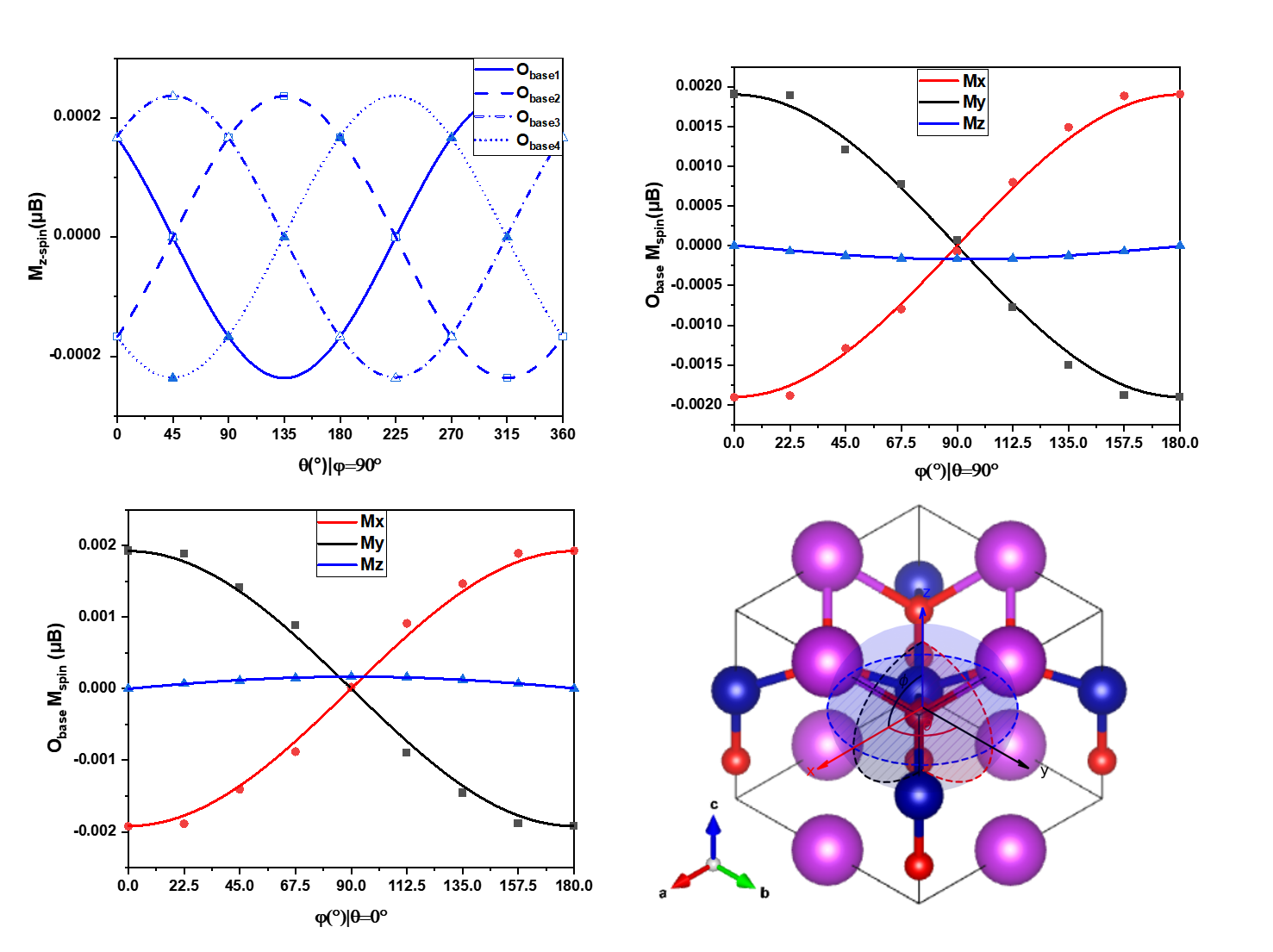}
    \caption{Behavior of the spin magnetization induced by a displacement of the basal oxygen atom, with fixed amplitude ($\tau$=0.03\r{A}), as a function of the direction of the displacement. The direction of the displacement is identified by polar coordinates $\theta$, $\phi$, as shown in the bottom right panel.}
    \label{fig:Oapibase}
\end{figure*}

\begin{figure}
    \includegraphics[width=0.6\linewidth]{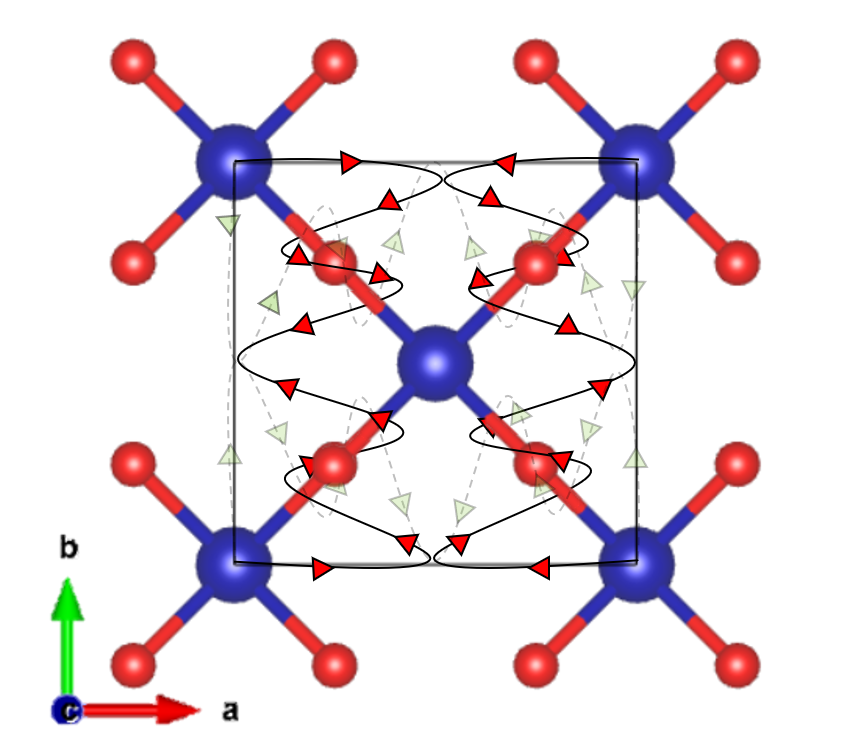}
    \caption{Sketch of the spiral behaviour along b of Obasal DMC created by the symmetry of the magnetic sub-lattices, the tendency is the same along a}
    \label{fig:SpiralDMCObasal}
\end{figure}

\begin{figure}
    \centering
    \includegraphics[width=0.5\linewidth]{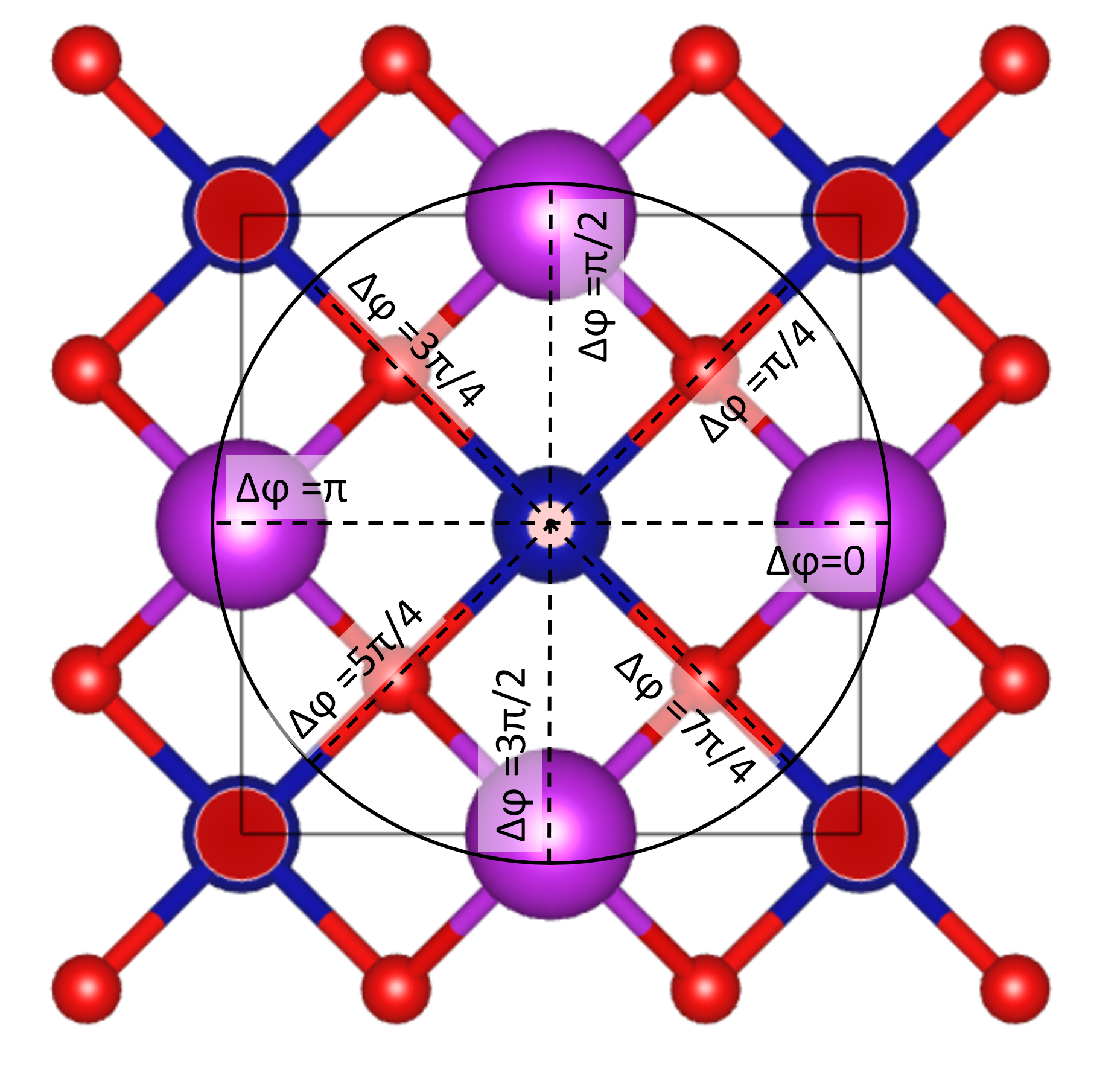}
    \caption{Gauge difference of the DMC between atoms and spin up cobalt whose gauge is fixed to zero}
    \label{fig:PhaseDiffAtoms}
\end{figure}
\clearpage

\section{Supplementary tables}

\begin{table*}[!hptb]
\caption{Behavior of spin dynamical magnetic charges (at a magnitude of $10^{-2}$ $\mu$B/\r{A}) atomic displacement.  $\theta$ is the azimuthal angle and $\phi$ the polar angle . Additionally, the parameter $\Delta\varphi$ denotes the gauge difference between considered atom and cobalt see Fig :~\ref{fig:PhaseDiffAtoms}  }
\begin{tabular}{c||c|c}
\hline\hline
\centering
Atom & ${\partial \Vec{S}_{x}(\mu B*10^{-2})/}{\partial r(\theta,\phi)}$  & ${\partial \Vec{S}_{y}(\mu B*10^{-2})/}{\partial r(\theta,\phi)}$                                         \\\hline

Bi    & $6.93 \cdot\sin{(\theta+\Delta\varphi)} \sin{(\phi)}$                                                           & $-6.93 \cdot\cos{(\theta+\Delta\varphi)} \sin{(\phi)}$                                                                                                                                                                                  \\\hline
Co    & $\sin{(\phi)}[6.53 \cdot\cos{(\theta)} $+$0.65 \cdot \sin{(\theta)}]$ & $\sin{(\phi)}[6.53 \cdot\sin{(\theta)}-0.65 \cdot \cos{(\theta)}$]                                                                                             \\\hline
O$_{api}$   & $2.80 \cdot\cos{(\theta)}\sin{(\phi)}$                                                            & $2.80 \cdot\sin{(\theta)}sin{(\phi)}$                                                                                                                                                    \\\hline
O$_{base}$  & $-5.77 \cdot\sin{(\theta+\Delta\varphi)}\cos{(\phi)}$                                                            & $5.77 \cdot\cos{(\theta+\Delta\varphi)}\cos{(\phi)}$               \\\hline                                            
\end{tabular}

\begin{tabular}{c||c}
\hline 
\centering
Atom & ${\partial \Vec{S}_{z}(\mu B*10^{-2})/}{\partial r(\theta,\phi)}$ \\\hline
Bi    & 0 \\\hline
Co   & $-0.91 \cdot \cos{(\phi)}$\\\hline
O$_{api}$  &  $-0.11 \cdot\cos{(\theta)}$ \\\hline
O$_{base}$  & $1.49 \cdot\cos{(\theta+\Delta\varphi)}\sin{(\phi)}+0.75 \cdot\sin{(\phi)}\cos{(2\theta+\Delta\varphi)}$
\\\hline 
\end{tabular}
\end{table*}

\renewcommand{\arraystretch}{1.3}
\begin{table*}[!hptb]
\caption{2$^{nd}$ order spin and orbital {\textit{diagonal}} dynamical magnetic effective charges (10$^{-2}\mu_{B}/\text{\r{A}}^2$) for each inequivalent Wyckoff sites of BCO as calculated with VASP and ABINIT (values in brackets in italic). $\mathcal{H}_i$ refers to the magnetic field in direction $i$ and $u_j$ to the atom displacement along the direction $j$.
}
\begin{tabular}{c|c|cccccclccccccc}
\hline\hline
    Atom          &     Wyc           & \multicolumn{7}{c}{$\Vec{S}$} & \multicolumn{7}{|c}{$\Vec{L}$} \rule{0pt}{1.2em}\\ 

\hline
                  &                   &  & \multicolumn{2}{c}{$\partial/\partial \mathcal{H}_{x}$} & \multicolumn{2}{c}{$\partial/\partial \mathcal{H}_{y}$} & \multicolumn{2}{c|}{$\partial/\partial \mathcal{H}_{z}$} &      & \multicolumn{2}{c}{$\partial/\partial \mathcal{H}_{x}$} & \multicolumn{2}{c}{$\partial/\partial \mathcal{H}_{y}$} & \multicolumn{2}{c}{$\partial/\partial \mathcal{H}_{z}$} \\ 
\multirow{3}{*}{Co}   & \multirow{3}{*}{1b}       & $\partial/\partial^2 u_{x}$  &  -     &   &   -    &       &   ()   & \multicolumn{1}{l|}{ } & $\partial/\partial^2 u_{x}$   &   -     &               &   -     &              & 83     &  ( \textit{186} )       \\ 
                  &                               & $\partial/\partial^2 u_{y}$  &  -     &   &   -    &       &   ()   & \multicolumn{1}{l|}{}  & $\partial/\partial^2 u_{y}$   &   -     &               &   -     &              &  83     &  ( \textit{186} )     \\
                  &                               & $\partial/\partial^2 u_{z}$  &  -     &   &   -    &       &  -4 (\textit{-4})  & \multicolumn{1}{l|}{}  & $\partial/\partial^2 u_{z}$   &   -     &               &   -     &              & 11      & (\textit{0})     \\ \hline
\multirow{3}{*}{O$_{api}$}  & \multirow{3}{*}{1b} & $\partial/\partial^2 u_{x}$  &  -     &   &   -    &       &   -    & \multicolumn{1}{l|}{}  & $\partial/\partial^2 u_{x}$    & -  &    & -  &     &   -   &     \\ 
                  &                               & $\partial/\partial^2 u_{y}$  &  -     &   &   -    &       &   -    & \multicolumn{1}{l|}{}  & $\partial/\partial^2 u_{y}$   & -  &    & -  &     &   -   &     \\
                  &                               & $\partial/\partial^2 u_{z}$  &  -     &   &   -    &       &   -    & \multicolumn{1}{l|}{}  & $\partial/\partial^2 u_{z}$   & -  &    & -  &     &   0   &   (\textit{2})  \\ \hline
\multirow{3}{*}{O$_{bas}$}  & \multirow{3}{*}{2c} & $\partial/\partial^2 u_{x}$  &  -     &  (\textit{-32})  &   -    &   -    &   -    & \multicolumn{1}{l|}{}  & $\partial/\partial^2 u_{x}$   &  -33 &  (\textit{34})  & -  &  - &  -         &        \\ 
                  &                               & $\partial/\partial^2 u_{y}$  &  0     &  -  &   0    &  (\textit{32})     &   -    & \multicolumn{1}{l|}{}  & $\partial/\partial^2 u_{y}$   & -  &    &  33  &  (\textit{34})  &   -       &          \\
                  &                               & $\partial/\partial^2 u_{z}$  &  -     &   &   -    &       &   -    & \multicolumn{1}{l|}{}  & $\partial/\partial^2 u_{z}$   &    -    &               & -       &              & -          &      \\ \hline
\end{tabular}
\label{tab:2ndordeu_sdmc_odmc}
\end{table*}
\renewcommand{\arraystretch}{1.0}
\bibliographystyle{unsrtnat} 
\bibliography{Submission/main}